\documentclass[3p]{elsarticle}

\usepackage{color}
\usepackage{amssymb}
\usepackage{graphicx}
\usepackage{amsmath}
\usepackage{subfig}
\usepackage{booktabs}
\usepackage{tikz}
\usepackage{pgfplots}
\usepgfplotslibrary{external}
\tikzset{external/system call={lualatex \tikzexternalcheckshellescape -halt-on-error -interaction=batchmode -jobname "\image" "\texsource"}}
\usepackage{placeins}
\usepackage{morefloats}

\journal{Journal of Computational Physics}

\begin{document}

\begin{frontmatter}

\title{Jacobian-free Newton-Krylov methods with GPU acceleration \\ for computing nonlinear ship wave patterns}

\author[Scott]{Ravindra Pethiyagoda}
\author[Scott]{Scott W. McCue\corref{cor1}}
\ead{scott.mccue@qut.edu.au}
\author[Scott]{Timothy J. Moroney}
\author[Scott]{Julian M. Back}

\cortext[cor1]{Corresponding author: Tel: +61 (0)7
31384295, Fax: +61 (0)7 31381508}

\address[Scott]{School of Mathematical Sciences, Queensland University of Technology, QLD 4101, Australia}

\begin{abstract}
The nonlinear problem of steady free-surface flow past a submerged source is considered as a case study for three-dimensional ship wave problems.  Of particular interest is the distinctive wedge-shaped wave pattern that forms on the surface of the fluid.  By reformulating the governing equations with a standard boundary-integral method, we derive a system of nonlinear algebraic equations that enforce a singular integro-differential equation at each midpoint on a two-dimensional mesh.  Our contribution is to solve the system of equations with a Jacobian-free Newton-Krylov method together with a banded preconditioner that is carefully constructed with entries taken from the Jacobian of the linearised problem.  Further, we are able to utilise graphics processing unit acceleration to significantly increase the grid refinement and decrease the run-time of our solutions in comparison to schemes that are presently employed in the literature.  Our approach provides opportunities to explore the nonlinear features of three-dimensional ship wave patterns, such as the shape of steep waves close to their limiting configuration, in a manner that has been possible in the two-dimensional analogue for some time.
\end{abstract}


\end{frontmatter}

\section{Introduction}
\label{sec:intro}

This study is concerned with steady three-dimensional free-surface profiles that are caused by a disturbance to a free stream.  These profiles are characterised by the distinctive Kelvin ship wave patterns that are observed at the stern of a vessel or even behind a duck swimming in an otherwise still body of water.  While free-surface flows of this type have ongoing practical applications to ship hull design, as we mention below, the structure of these patterns has sparked renewed interest in the physics literature, with observations that ships moving sufficiently fast may give rise to wake angles that decrease with ship speed~\cite{darmon14,ellingsen14,rabaud13}, in apparent contradiction to the well-known Kelvin angle of $\arcsin (1/3)\approx 19.47^\circ$ \cite{lighthill}, which is derived from linear theory.  In contrast to these approaches, our purpose here is to treat the fully {\em nonlinear} equations, and present algorithms for the accurate computation of nonlinear ship wave profiles.

The mathematical analysis of ship wave patterns has a very long history, the overwhelming majority of which concerns linear theories.  For example, for the classic problem of flow past a pressure distribution applied to the surface of the fluid $z=\zeta(x,y)$, if the pressure is small enough then the kinematic and Bernoulli boundary conditions on $z=\zeta(x,y)$ can be linearised onto the undisturbed plane $z=0$ \cite{havelock19,noblesse09b,scullen11}.  This framework is used to model the wave pattern caused by an air-cushioned vehicle such as a hovercraft or a high-speed ``flat ship'' with a small draft.  Another approach is to consider the ship wave pattern due to a thin ship.  In this case the no-flux conditions on the ship hull are linearised onto the centreplane $y=0$, while the thinness of the ship is assumed to produce small-amplitude waves, so the free surface conditions are again linearised onto the plane $z=0$ \cite{michell98,noblesse09a,tuck01}.  This set-up has obvious applications to ship hull design, especially for vessels with narrow hulls.  Another geometry of interest involves flow past a submerged object, such as a spheroid, or, in a fluid of finite-depth, a bottom topography.  If the magnitude of the disturbance is again small, then the usual linearisation of the surface conditions applies \cite{havelock31}.  Furthermore, one can apply the thin ship approximation to submerged bodies as well \cite{tuck02}.  Flows past submerged bodies have applications to submarine design and detection \cite{reed02}, for example.

In all of the linear formulations cited above, the linear problem of Laplace's equation in a known domain can be solved in principle with Fourier transforms \cite{lighthill}.  The velocity potential $\phi(x,y,z)$ and free surface $z=\zeta(x,y)$ are then given as quadruple integrals that involve the Havelock potential (the Green's function or fundamental solution).  In practice, the challenge of evaluating the resulting singular integrals with rapidly oscillating integrands has lead to analytical approximations such as the method of stationary phase \cite{crapper64,tuck71,ursell60}, although accurate numerical computations have been conducted more recently \cite{noblesse09a,noblesse09b,scullen11}.  Of particular interest here, we note that the Havelock potential is the velocity potential for the linearised flow past a single submerged point source singularity \cite{havelock32,lustrichapman13,noblesse78,noblesse81,peters49}.  Thus we see that the problem of computing the wave pattern caused by turning on a submerged source in a uniform stream acts as a building block for all the other flows mentioned (as an example, the thin-ship theory effectively states that the flow past a thin ship hull is equivalent to the flow past a distribution of point sources on the centreplane $y=0$ whose strength is proportional to the hull slope $\partial y/\partial x$ \cite{tuck01}).

Our focus in this study is to compute nonlinear flows, for which the full nonlinear boundary conditions on the actual displaced free surface $z=\zeta(x,y)$ apply.  Nonlinear versions of the above problems have been considered by a number of authors (see \cite{higgins12,parau02,tuck02}, for example).  In particular, following the framework of Forbes~\cite{forbes89}, the approach we are most interested in is to apply a boundary-integral technique that relies on Green's second formula.  The result is a singular integro-differential equation which holds on the unknown free surface $z=\zeta(x,y)$.  That is, the free-surface problem in three dimensions is reduced to a two-dimensional problem for the free surface $z=\zeta(x,y)$ and the velocity potential $\phi(x,y,\zeta(x,y))$.  To proceed numerically, the rough approach is to place a mesh of $N\times M$ grid points over the truncated $(x,y)$-plane, so that the integro-differential equation and Bernoulli's equation can both be applied at each of the $(N-1)M$ half-mesh points. A radiation-type condition for the four unknown functions is applied at each of the $M$ grid points upstream. Newton's method is then used to solve the resulting nonlinear system of $2(N+1)M$ equations for the $2(N+1)M$ unknowns (which are slopes $\partial\zeta/\partial x$, $\partial\phi/\partial x$ and the values of $\zeta$, $\phi$ on the upstream grid points).  As discussed by Forbes~\cite{forbes89}, moderate efficiencies can be gained by exploiting the symmetry of the problem and using an inexact Newton's method which re-uses the Jacobian a number of times if possible.

In more recent times, over a series of papers, P\u{a}r\u{a}u, Vanden-Broeck and Cooker have applied Forbes' formulation to solve fully three-dimensional nonlinear ship wave problems and have typically used meshes of between $60\times 20$ and $80\times 40$ grid points \cite{parau02,parau11,parau07b,parau07c,parau07a,parau10}.  The same authors applied the same formulation to study three-dimensional solitary waves with typical meshes of $50\times 40$ grid points \cite{parau05a,parau05b,parau11}, while Forbes \& Hocking~\cite{forbes05} used a mesh of $101\times 101$ points when applying the method to a three-dimensional withdrawal problem.  To put the method into context, other approaches for three-dimensional ship wave problems use a similar grid size; for example, Tuck \& Scullen~\cite{tuck02} apply a mesh of $91\times 25$ grid points with their Rankine source method, while similar resolution is provided for a Rankine source method in \cite{Janson03}.

The level of grid refinement demonstrated for the three-dimensional problems just mentioned is to be contrasted with the vast literature on two-dimensional flows.  For example, by applying a boundary-integral method in two dimensions combined with a straight-forward Newton approach, authors can easily use in excess of 1000 grid points over the two-dimensional surface \cite{mccue02,ogilat11,wade14}
or, in more recent times, even 2000 points \cite{lustri12,trinh14,trinh11}.
Although most authors end up using fewer than 1000 points for their two-dimensional calculations, generally an accepted procedure is to continue to refine the mesh until the results are grid-independent, at least visually.  Turning our attention back to three-dimensional flows, with less than 100 points used along the $x$-direction, the resolution over each wavelength is simply not of a sufficient standard for any claims about grid-independence to be made.  Indeed, this is one of the key reasons why there has been little to no detailed study of the effect of high nonlinearity for three-dimensional ship wave problems.

In the present paper, we use a variation of the numerical scheme developed by Forbes~\cite{forbes89} for the problem of flow past a submerged source singularity, and apply Jacobian-free Newton-Krylov methods and exploit graphics processing unit (GPU) acceleration to drastically increase the grid refinement and decrease the run-time when compared with schemes published in the literature.  We choose this particular geometric configuration since, as mentioned above, it can be thought of as the most fundamental flow type within the class that produces three-dimensional ship wave patterns.  Further, this is precisely the geometry that Forbes~\cite{forbes89} used when presenting the boundary-integral technique described above.  Thus we have a direct correspondence and a bigger picture view of how far the community has progressed since that time.  Finally, all of our ideas should generalise for other configurations (such as flows past pressure distributions), provided there is a linear problem that arises in the small disturbance regime.

In the following section we formulate the problem of interest and provide a summary of the boundary-integral technique developed by Forbes~\cite{forbes89} and P\u{a}r\u{a}u and Vanden-Broeck~\cite{parau02}.  The numerical scheme is described in Section~\ref{sec:numerical}, which leads to a nonlinear system of equations
\begin{equation}
{\bf E}({\bf u})={\bf 0},
\label{eq:nonlinearsys}
\end{equation}
where ${\bf u}$ is the vector of unknowns of length $2(N+1)M$.  The damped Newton's method approach leads to the iteration
\begin{equation}
\textbf{u}_{k+1} = \textbf{u}_{k} + \lambda_k \delta\textbf{u}_k,
\label{eq:newtonstep0}
\end{equation}
where $\textbf{u}_{k}$ is the $k$th iterate in the sequence $\{ {\bf u}_k \}_{k=0}^{\infty}\rightarrow {\bf u}$ and the damping parameter $\lambda_k\in(0,1]$ is chosen such that $||\textbf{E}(\textbf{u}_{k+1})||<||\textbf{E}(\textbf{u}_{k})||$ at every iterate.  The Newton step $\delta\textbf{u}_k$ satisfies
\begin{equation}
\textbf{J}(\textbf{u}_k) \delta\textbf{u}_k = -\textbf{E}(\textbf{u}_k),
\label{eq:newtonstep}
\end{equation}
where ${\bf J}=\partial {\bf E}/\partial {\bf u}$ is the Jacobian matrix.  The integral nature of our governing equations results in  all of the entries in ${\bf u}$ contributing to the evaluation of each component of $\bf E$ that corresponds to enforcing the integral equation, which means that the lower-half of the Jacobian ${\bf J}$ is fully dense.  This density has been a significant factor in limiting the number of grid points used in previously published numerical simulations.

A key aspect of our approach is the use of a Jacobian-free Newton-Krylov method to solve the system (\ref{eq:nonlinearsys}).  A Jacobian-free Newton-Krylov method requires the action of the Jacobian only in the form of Jacobian-vector products, which can be approximated using difference quotients without ever forming the Jacobian itself \cite{knoll04}.  In practice, the underlying Krylov subspace iterative solver requires preconditioning in order to achieve a satisfactory rate of convergence, meaning the overall method is not typically fully matrix-free; however, for preconditioning purposes, an approximation of the Jacobian is all that is required, and this is where significant savings can be made.

While Jacobian-free Newton-Krylov methods are most commonly associated with problems for which the Jacobian matrices are sparse, they have been used successfully in a number of applications that give rise to dense Jacobian matrices \cite{chacon00, khatiwala08, moroney13}.  In each of these applications, a sparse approximation of the Jacobian was used in constructing the preconditioner.  We take the same approach in this work.  The type of approximation we find to be the most effective involves a banded structure, with its nonzero entries coming from the linearised problem for a Havelock source mentioned above.  We emphasise that this approximation is used only for preconditioning purposes; the action of the dense Jacobian is still felt throughout the Newton solver, which distinguishes our approach from the inexact method of Forbes~\cite{forbes89} and others.

In Section~\ref{sec:results} we present our results.  We choose to present most of our results for a particular set of parameter values, which includes the same Froude number as used by Forbes~\cite{forbes89}, and a moderately large value of the dimensionless strength of the submerged source.  While Forbes showed results computed with a mesh of $45\times 13$ grid points in 1989, we are able to easily use a $361\times 121$ mesh on a modern desktop PC, computed in under 75 minutes.  By utilising graphics processing unit (GPU) acceleration on a more powerful workstation, the same solution was computed in roughly 3.5 minutes.  Furthermore, with this technology we are able to significantly improve upon the resolution, and generate results for a $721\times 241$ mesh (in under 2 hours).  This sort of resolution is important for three-dimensional ship wave problems, as it provides opportunities to explore the effect that nonlinearity has on the flow field in the same way as has been done in numerous instances for two-dimensional flows.  Finally, we close the paper in Section~\ref{sec:discussion} with our discussion, including directions as to where our work can be applied.


\section{Mathematical formulation}
\label{sec:formulation}

\subsection{Governing equations}

We consider the irrotational flow of an inviscid, incompressible fluid of infinite depth, bounded above by a free surface, upon which gravity is acting.  The effects of surface tension are ignored.  Suppose that initially there is a free stream of fluid travelling with uniform speed $U$ in the positive $x$-direction, and that a source singularity of strength $m$ is introduced at a distance $H$ below the surface.  The disturbance caused by the source will lead to transient waves being generated on the free-surface.  We are interested in the steady-state problem that arises in the long-time limit of this flow.

The problem is nondimensionalised by scaling all lengths with respect to $H$ and all speeds with respect to $U$.  By labelling the free-surface $z=\zeta(x,y)$, the dimensionless problem is to solve Laplace's equation for the velocity potential $\Phi(x,y,z)$:
\begin{align}
\nabla^2\Phi=\frac{\partial^2 \Phi}{\partial x^2}+\frac{\partial^2 \Phi}{\partial y^2}+\frac{\partial^2 \Phi}{\partial z^2}=0 \quad \text{for} \quad z<\zeta(x,y),\label{eqn:NondimLap}
\end{align}
except at the source singularity itself, whose dimensionless location is at $(x,y,z)=(0,0,-1)$.  The appropriate limiting behaviour is
\begin{align}
\Phi\sim-\frac{\epsilon}{4\pi\sqrt{x^2+y^2+\left(z+1\right)^2}} \quad \text{as} \quad (x,y,z)\rightarrow(0,0,-1),\label{eqn:NondimSource}
\end{align}
where
\begin{equation}
\epsilon=\frac{m}{UH^2}\label{eqn:epsilon}
\end{equation}
is the dimensionless source strength.  On the free surface there are the kinematic and dynamic boundary conditions
\begin{align}
\Phi_x\zeta_x+\Phi_y\zeta_y&=\Phi_z \quad \text{on} \quad z=\zeta(x,y),\label{eqn:NondimKin}\\
\frac{1}{2}(\Phi^2_x+\Phi^2_y+\Phi^2_z)+\frac{\zeta}{F^2}&=\frac{1}{2} \quad\;\: \text{on} \quad z=\zeta(x,y),\label{eqn:NondimDyn}
\end{align}
being satisfied, where the second of two dimensionless parameters in the problem is the depth-based Froude number
\begin{equation}
F=\frac{U}{\sqrt{gH}}.\label{eqn:Froude}
\end{equation}
Finally, the flow will approach the free stream both far upstream (the radiation condition) and infinitely far below the free surface, providing the final two conditions
\begin{align}
(\Phi_x,\Phi_y,\Phi_z)\rightarrow(1,0,0),&\quad \zeta \rightarrow 0\quad\text{as} \quad x\rightarrow-\infty,\label{eqn:NondimUp}
\\
(\Phi_x,\Phi_y,\Phi_z)\rightarrow(1,0,0),&\qquad\qquad\;\;\text{as}\quad z\rightarrow-\infty.\label{eqn:NondimFar}
\end{align}
The governing equation (\ref{eqn:NondimLap}) subject to (\ref{eqn:NondimSource})-(\ref{eqn:NondimFar}) make up a nonlinear free-surface problem with no known analytical solution.

\subsection{Boundary-integral method}

In order to solve (\ref{eqn:NondimLap})-(\ref{eqn:NondimFar}) numerically, we first reformulate the problem in terms of an integral equation using Green's second formula.  The full derivation is provided in Forbes~\cite{forbes89}, while very similar approaches are outlined in a variety of other papers \cite{forbes05,parau02,parau11,parau05a,parau05b,parau07b,parau07c,parau07a,parau10}.
By setting $\phi(x,y)=\Phi(x,y,\zeta(x,y))$, the final boundary-integral equation is
\begin{align}
2\pi(\phi(x^*,y^*)-x^*)=&-\frac{\epsilon}{\left({x^*}^2+{y^*}^2+(\zeta(x^*,y^*)+1)^2 \right)^\frac{1}{2}}\notag\\
&+\int\limits_{0}^{\infty}\int\limits_{-\infty}^{\infty}(\phi(x,y)-\phi(x^*,y^*)-x+x^*)K_1(x,y;x^*,y^*)\,\, \text{d}x\,\text{d}y\notag\\
&+\int\limits_{0}^{\infty}\int\limits_{-\infty}^{\infty}\zeta_x(x,y)K_2(x,y;x^*,y^*)\,\, \text{d}x\,\text{d}y,\label{eqn:IntegroEqn}
\end{align}
which holds for any point $(x^*,y^*)$ in the $(x,y)$-plane.  Here $K_1$ and $K_2$ are the kernel functions
\begin{align*}
K_1(x,y;x^*,y^*)=&\frac{\zeta(x,y)-\zeta(x^*,y^*)-(x-x^*)\zeta_x-(y-y^*)\zeta_y}{\Bigl((x-x^*)^2+(y-y^*)^2+\bigl(\zeta(x,y)-\zeta(x^*,y^*)\bigr)^2\Bigr)^\frac{3}{2}}\\
&+\frac{\zeta(x,y)-\zeta(x^*,y^*)-(x-x^*)\zeta_x-(y+y^*)\zeta_y}{\Bigl((x-x^*)^2+(y-y^*)^2+\bigl(\zeta(x,y)-\zeta(x^*,y^*)\bigr)^2\Bigr)^\frac{3}{2}},\\
K_2(x,y;x^*,y^*)=&\frac{1}{\sqrt{(x-x^*)^2+(y-y^*)^2+\bigl(\zeta(x,y)-\zeta(x^*,y^*)\bigr)^2}}\\ &+\frac{1}{\sqrt{(x-x^*)^2+(y+y^*)^2+\bigl(\zeta(x,y)-\zeta(x^*,y^*)\bigr)^2}}.
\end{align*}

The integral equation (\ref{eqn:IntegroEqn}) identically satisfies Laplace's equation (\ref{eqn:NondimLap}) and the kinematic condition (\ref{eqn:NondimKin}), as well as the limiting condition (\ref{eqn:NondimSource}) and the far-field conditions (\ref{eqn:NondimUp})-(\ref{eqn:NondimFar}).  Thus we are left to solve (\ref{eqn:IntegroEqn}) and the dynamic condition (\ref{eqn:NondimDyn}).  It proves convenient to rewrite (\ref{eqn:NondimDyn}) with the help of (\ref{eqn:NondimKin}) to be
\begin{align}
\frac{1}{2}\frac{(1+\zeta_x^2)\phi_y^2+(1+\zeta_y^2)\phi_x^2-2\zeta_x\zeta_y\phi_x\phi_y}{1+\zeta_x^2+\zeta_y^2}+\frac{\zeta}{F^2}=\frac{1}{2},
\quad &\text{on} \quad z=\zeta(x,y).
\label{eqn:freeSurfCond}
\end{align}

\subsection{Linearised problem}\label{sec:linearproblem}

While our focus is on generating numerical solutions to (\ref{eqn:NondimLap})-(\ref{eqn:NondimFar}), it will prove instructive to note the linearised problem which arises in the weak source strength limit $\epsilon\ll 1$.  The problem is formulated by writing $\Phi=x+\epsilon \Phi_1(x,y,z)+\mathcal{O}(\epsilon^2)$, $\zeta=\epsilon\zeta_1(x,y)+\mathcal{O}(\epsilon^2)$, and considering the formal limit $\epsilon\rightarrow 0$.  As a result, the linear problem becomes
\begin{align}
\nabla^2\Phi=\frac{\partial^2 \Phi}{\partial x^2}+\frac{\partial^2 \Phi}{\partial y^2}+\frac{\partial^2 \Phi}{\partial z^2}=0 \quad \text{for} \quad z<0,\label{eqn:NondimLaplinear}
\end{align}
subject to the linearised kinematic and dynamic conditions
\begin{align}
\zeta_x&=\Phi_z \quad \text{on} \quad z=0,\label{eqn:NondimKinlinear}\\
\Phi_x-1+\frac{\zeta}{F^2}&=0 \quad\;\: \text{on} \quad z=0.\label{eqn:NondimDynlinear}
\end{align}
The near-source behaviour (\ref{eqn:NondimSource}) and the far-field conditions (\ref{eqn:NondimUp})-(\ref{eqn:NondimFar}) remain the same.

As discussed in the Introduction, the solution to this linear problem can be found using Fourier transforms \cite{noblesse81}; however, for our purposes we shall pursue the equivalent boundary-integral approach as that used for the nonlinear problem.  This time if we set $\phi(x,y)=\Phi(x,y,0)$, the application of Green's second formula gives
\begin{equation}
2\pi(\phi(x^*,y^*)-x^*)=
-\frac{\epsilon}{\left({x^*}^2+{y^*}^2+1\right)^\frac{1}{2}}
+\int\limits_{0}^{\infty}\int\limits_{-\infty}^{\infty}\zeta_x(x,y)K_3(x,y;x^*,y^*)\,\, \text{d}x\,\text{d}y,\label{eqn:IntegroEqnlinear}
\end{equation}
where
\begin{equation*}
K_3(x,y;x^*,y^*)=\frac{1}{\sqrt{(x-x^*)^2+(y-y^*)^2}}
+\frac{1}{\sqrt{(x-x^*)^2+(y+y^*)^2}}.
\end{equation*}
Again, the integral equation (\ref{eqn:IntegroEqnlinear}) identically satisfies Laplace's equation (\ref{eqn:NondimLaplinear}), the kinematic condition (\ref{eqn:NondimKinlinear}), the far-field conditions (\ref{eqn:NondimUp})-(\ref{eqn:NondimFar}) and the near-source condition (\ref{eqn:NondimSource}).

\section{Numerical discretisation}
\label{sec:numerical}
For the discretisation of the nonlinear boundary-integral equation (\ref{eqn:IntegroEqn}), we use a slight variant of the method outlined in P\u{a}r\u{a}u and Vanden-Broeck~\cite{parau02}, which is based on the original approach of Forbes~\cite{forbes89}. This involves laying a regular mesh of nodes $(x_1, y_1), \ldots, (x_N, y_M)$ on the free surface with spacings of $\Delta x$ and $\Delta y$ in the $x$ and $y$ directions, respectively.  For a given $N$ and $M$, we shall refer to the mesh as being an $N\times M$ mesh.
The free-surface position $\zeta(x,y)$ and the velocity potential $\phi(x,y)$ are represented by discrete values $\zeta_{k,\ell}$ and $\phi_{k,\ell}$ at the points $(x_k,y_\ell),\, k=1,\dots,N,\, \ell=1,\dots,M$.

We define the vector of $2(N+1)M$ unknowns $\textbf{u}$ to be
\begin{align}
\textbf{u}=&[\phi_{1,1},(\phi_x)_{1,1},\dots,(\phi_x)_{N,1},\phi_{1,2},(\phi_x)_{1,2},\dots,(\phi_x)_{N,2},\ldots,\phi_{1,M}, (\phi_x)_{1,M},\dots,(\phi_x)_{N,M},
\nonumber \\
&\zeta_{1,1},(\zeta_x)_{1,1},\dots,(\zeta_x)_{N,1},\zeta_{1,2},(\zeta_x)_{1,2},\dots,(\zeta_x)_{N,2},\ldots, \zeta_{1,M},(\zeta_x)_{1,M},\dots,(\zeta_x)_{N,M}]^T,
\label{eq:unknowns}
\end{align}
comprising the $x$-derivatives of the functions $\phi$ and $\zeta$ at the free-surface mesh points, together with the values of $\phi$ and $\zeta$ at the upstream boundary of the truncated domain.  The values of these unknowns are related via $2(N+1)M$ nonlinear equations, of the form (\ref{eq:nonlinearsys}), which we now derive.

Given the elements of the vector of unknowns (\ref{eq:unknowns}), the remaining values of $\zeta$ are obtained by trapezoidal-rule integration using the values of $\zeta_x$:
\begin{equation}
	\begin{aligned}
	\zeta_{k+1,\ell}&=\zeta_{k,\ell}+\frac{1}{2}\Delta x\bigl((\zeta_x)_{k,\ell}+(\zeta_x)_{k+1,\ell} \bigr),\\
	\ell&=1,\dots,M,\quad k=1,\dots,N-1.
	\end{aligned}\label{eqn:zetaapprox}
\end{equation}
The values of $\zeta_y$ are then computed by fitting a cubic spline through the points $\zeta_{k,1},\dots,\zeta_{k,M}$ for $k=1,\dots,N$. Values of $\phi$ and $\phi_y$ at each grid point are similarly computed using $\phi_x$.

We must now enforce the integro-differential equation (\ref{eqn:IntegroEqn}), which will be evaluated on the half-mesh points $(x_{k+\frac{1}{2}},y_\ell),\, k=1,\dots,N-1,\, \ell=1,\dots,M$ using two-point interpolation. The domain is truncated to the rectangle $x_1\leq x\leq x_N,\ y_1\leq y\leq y_M$. The singularity in the second integral of (\ref{eqn:IntegroEqn}) is removed by the addition and subtraction of the term
\begin{equation}
\zeta_x(x^*,y^*)\int\limits_{y_1}^{y_M}\int\limits_{x_1}^{x_N}S_2(x,y;x^*,y^*)\,\, \text{d}x\,\text{d}y, \label{eqn:I2DashDash}
\end{equation}
where
\begin{align}
S_2(x,y;x^*,y^*)=&\frac{1}{\sqrt{A(x-x^*)^2+B(x-x^*)(y-y^*)+C(y-y^*)^2}}\\
&+\frac{1}{\sqrt{A(x-x^*)^2-B(x-x^*)(y+y^*)+C(y+y^*)^2}},\label{eqn:S2}
\end{align}
with
\begin{equation*}
A=1+\zeta_x^2(x^*,y^*),\quad B=2\zeta_x(x^*,y^*)\zeta_y(x^*,y^*),\quad C=1+\zeta_y^2(x^*,y^*).
\end{equation*}
The second integral of the equation (\ref{eqn:IntegroEqn}) becomes
\begin{equation*}
\int\limits_{y_1}^{y_M}\int\limits_{x_1}^{x_N}\zeta_x(x,y) K_2(x,y;x^*,y^*)-\zeta_x(x^*,y^*)S_2(x,y;x^*,y^*)\,\, \text{d}x\,\text{d}y +\zeta_x(x^*,y^*)I,
\end{equation*}
where
\begin{equation}
I=\int\limits_{y_1}^{y_M}\int\limits_{x_1}^{x_N}S_2\,\, \text{d}x\,\text{d}y.\label{eqn:IS2}
\end{equation}
The integral $I$ now contains the singularity; it can be evaluated exactly in terms of logarithms \cite{forbes89,parau02}.

The integrals in the approximation to equation (\ref{eqn:IntegroEqn}) are discretised using the trapezoidal rule and then evaluated for all half-mesh points $(x_{k+\frac{1}{2}},y_\ell),\, k=1,\dots,N-1,\, \ell=1,\dots,M$. This results in $(N-1)M$ nonlinear algebraic equations for the unknowns in the vector $\textbf{u}$.  An additional $(N-1)M$ equations are given by evaluating the free surface condition (\ref{eqn:freeSurfCond}) at the half mesh points. The final $4M$ equations are provided to enforce the far-field condition (\ref{eqn:NondimUp}) on the relevant boundary of the truncated domain by applying the upstream radiation condition using the approach outlined by Scullen~\cite{scullen98}.  The idea here is to enforce an equation of the form $xf_x+n f=0$ along the boundary $x=x_1$ for the four functions $\zeta$, $\zeta_x$, $\phi-x$ and $\phi_x-1$.  The value of $n>0$ represents how fast the functions decay to zero upstream, and in our calculation was taken to be $n=0.05$ (larger values of $n$ were found to amplify the small spurious upstream waves mentioned below).   This method for applying the radiation condition gives us the $4M$ equations
\begin{equation}
	\begin{aligned}
	x_1((\phi_x)_{1,\ell}-1)+n(\phi_{1,\ell}-x_1)&=0,\\
	x_1(\phi_{xx})_{1,\ell}+n((\phi_x)_{1,\ell}-1)&=0,\\
	x_1(\zeta_x)_{1,\ell}+n\zeta_{1,\ell}&=0,\\
	x_1(\zeta_{xx})_{1,\ell}+n(\zeta_x)_{1,\ell}&=0,
	\end{aligned}\label{eqn:numUpRadiation}
\end{equation}
for $\ell=1,\dots,M$, where second derivatives are computed by a forward difference approximation on the first derivative.  We now have $2(N+1)M$ equations for our vector of unknowns (\ref{eq:unknowns}).  In order to optimise our scheme we have ordered these equations very carefully.  This ordering is explained in \ref{appendixA}.

This numerical scheme has two main sources of error. The first is truncation error introduced when approximating the infinite domain of integration with a finite domain.  This truncation has the potential to lead to errors if the chosen upstream truncation point ($x_1$) is too close to the source, as the upstream radiation condition (\ref{eqn:numUpRadiation}) may no longer be accurately enforced. Indeed, truncating the domain upstream appears to generate very small nonphysical waves on the surface, as discussed later. Truncating the domain downstream (at $x_N$) may also introduce significant errors as the amplitude of the wavetrain decays slowly with space, and contribution to the integrals from the truncated waves is nonzero.  The second main source of error is from the discretisation of the integrals. Both the mesh spacing and the chosen integration weighting scheme will have an effect on the accuracy of the final result.

\section{Jacobian-free Newton-Krylov method}

\subsection{Overview}

The system (\ref{eq:nonlinearsys}) is solved with a Jacobian-free Newton-Krylov method.  At the outer, nonlinear level, this is simply the damped Newton iteration (\ref{eq:newtonstep0}), with $\lambda_k$ chosen via a simple linesearch to ensure a sufficient decrease in the nonlinear residual is obtained with each iteration. At the inner, linear level, the system (\ref{eq:newtonstep}) is solved using the iterative Generalised Minimum Residual algorithm~\cite{saad86} with right preconditioning.  After $m$ iterations of this algorithm, the approximate solution for the Newton correction $\delta \mathbf{u}_k$ is found by projecting obliquely onto the preconditioned Krylov subspace
\[
\mathcal{K}_m(\textbf{J}_k\mathbf{P}^{-1},\textbf{E}_k)=\mathrm{span}\{\textbf{E}_k,\textbf{J}_k\mathbf{P}^{-1}\textbf{E}_k,\dots,(\textbf{J}_k\mathbf{P}^{-1})^{m-1}\textbf{E}_k\},
\]
where we are now using the notation $\textbf{J}_k=\textbf{J}(\textbf{u}_k)$, $\textbf{E}_k=\textbf{E}(\textbf{u}_k)$.  The matrix $\mathbf{P} \approx \mathbf{J}_k$ is the preconditioner matrix -- a sparse approximation to $\textbf{J}_k$ which is discussed in more detail in the next subsection.  Its function is to reduce the dimension $m$ of the Krylov subspace required to obtain a sufficiently accurate solution for $\delta \mathbf{u}_k$.

Krylov subspace methods are very attractive as linear solvers in the context of nonlinear Newton iteration, because they do not require explicit formation of the Jacobian matrix.  Indeed, only the action of the Jacobian matrix in the form of Jacobian-vector products is required to assemble a basis for the preconditioned Krylov subspace $\mathcal{K}_m$.  These Jacobian-vector products can be approximated without needing to form $\textbf{J}_k$ by using first order difference quotients:
\begin{equation}
\textbf{J}_k\mathbf{P}^{-1}\textbf{v} \approx \frac{\textbf{E}(\textbf{u}_k+h\,\mathbf{P}^{-1}\textbf{v})-\textbf{E}(\textbf{u}_k)}{h},\label{eqn:JvApprox}
\end{equation}
where $\textbf{v}$ represents an arbitrary vector used in building the Krylov subspace, and $h$ is a suitably-chosen shift \cite{brown90}.

Since the Newton correction is solved for only approximately, and the action of the Jacobian in computing this solution is itself only approximated, we are left with an inexact Newton method, which exhibits superlinear, rather than quadratic, convergence \cite{knoll04}.  The reduction in the convergence rate is of little practical consequence, given the enormous performance gains realised by removing the burden of forming the (dense) Jacobian matrix.  Furthermore, only solving for the Newton correction approximately can actually improve performance in the early stages of the nonlinear iteration, by not wasting operations computing an extremely accurate value of the Newton correction which, even if it were computed exactly, would only reduce the nonlinear residual by so much \cite{knoll04}.

For most values of the parameters $F$ and $\epsilon$, it proves sufficient to use a flat surface as the initial guess ${\bf u}_0$ in the Newton iteration, which corresponds to:
\begin{equation*}
\phi_{1,\ell}=x_0,
\quad
(\phi_x)_{k,\ell}=1,
\quad
\zeta_{1,\ell}=0,
\quad
(\zeta_x)_{k,\ell}=0,
\end{equation*}
for $k=1,\dots, N$ and $\ell=1,\dots, M$. Another approach is to use the exact solution to the linear problem outlined in Section~\ref{sec:linearproblem} (given in \cite{noblesse81}, for example).  However, for highly nonlinear solutions with large values of $\epsilon$, a further alternative approach is to apply a bootstrapping process in which a solution is computed using ${\bf u}_0$ for a moderate value of $\epsilon$, and then this solution is used as an initial guess for a slightly larger $\epsilon$, and so on.

\subsection{Preconditioning}\label{sec:preconditioning}
In forming the preconditioner matrix $\mathbf{P}$, the goal is to construct an approximation to the Jacobian $\mathbf{J}_k$ that is cheap to form and to factorise, such that the spectrum of the preconditioned Jacobian $\mathbf{J}_k\mathbf{P}^{-1}$ exhibits a clustering of eigenvalues \cite{knoll04}.  A common starting point in building such a preconditioner is to consider a matrix constructed from the same problem under simplified physics~\cite{knoll04}.  In the present context, this is achieved by applying our numerical scheme to the linearised governing equations which apply formally in the limit $\epsilon\rightarrow 0$.  These equations make up the well-studied linear problem of computing the Havelock potential for flow past a submerged point source \cite{havelock32,lustrichapman13,noblesse78,noblesse81,peters49}, as discussed in the Introduction and Section~\ref{sec:linearproblem}.
The numerical discretisation of the integrals in (\ref{eqn:IntegroEqnlinear}) allows for easy differentiation by hand, so that all elements of the linear Jacobian can be calculated exactly, requiring considerably less computational time. The details are included in \ref{appendixB}.
\begin{figure}[h]
\centering
\subfloat[Nonlinear Jacobian]{\label{fig:JacVis}\includegraphics[width=.45\linewidth]{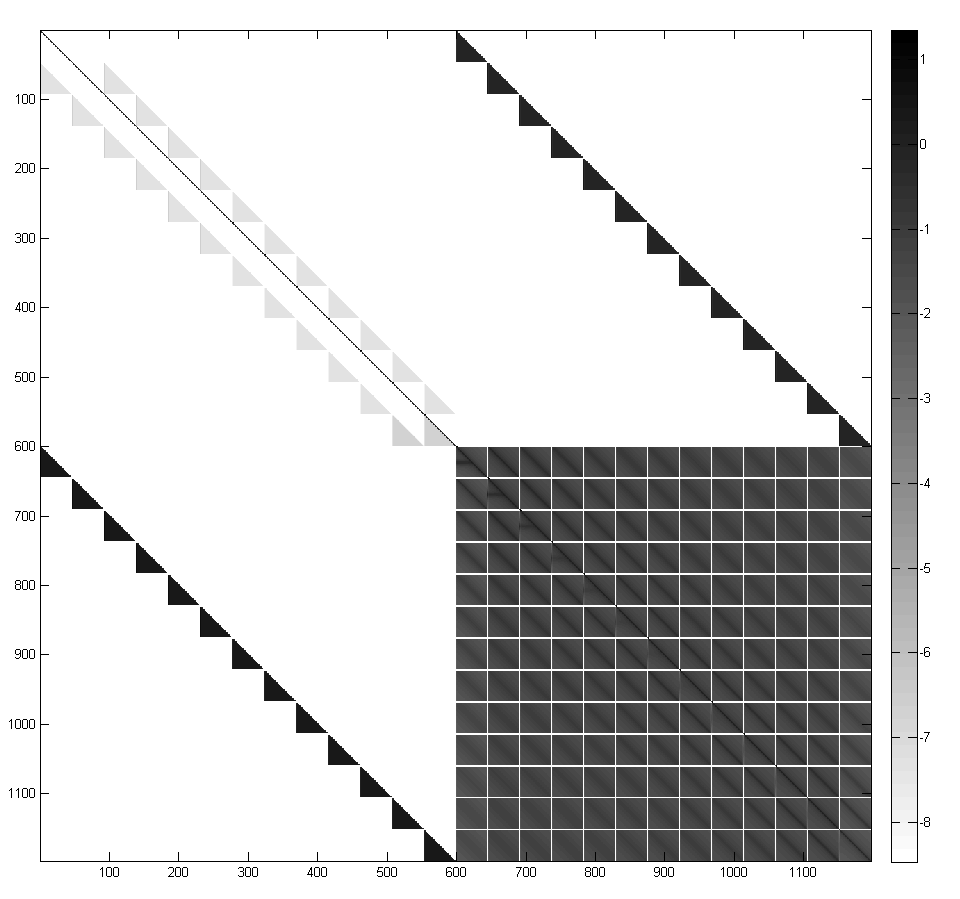}}
\subfloat[Linear Jacobian]{\label{fig:JacLinVis}\includegraphics[width=.45\linewidth]{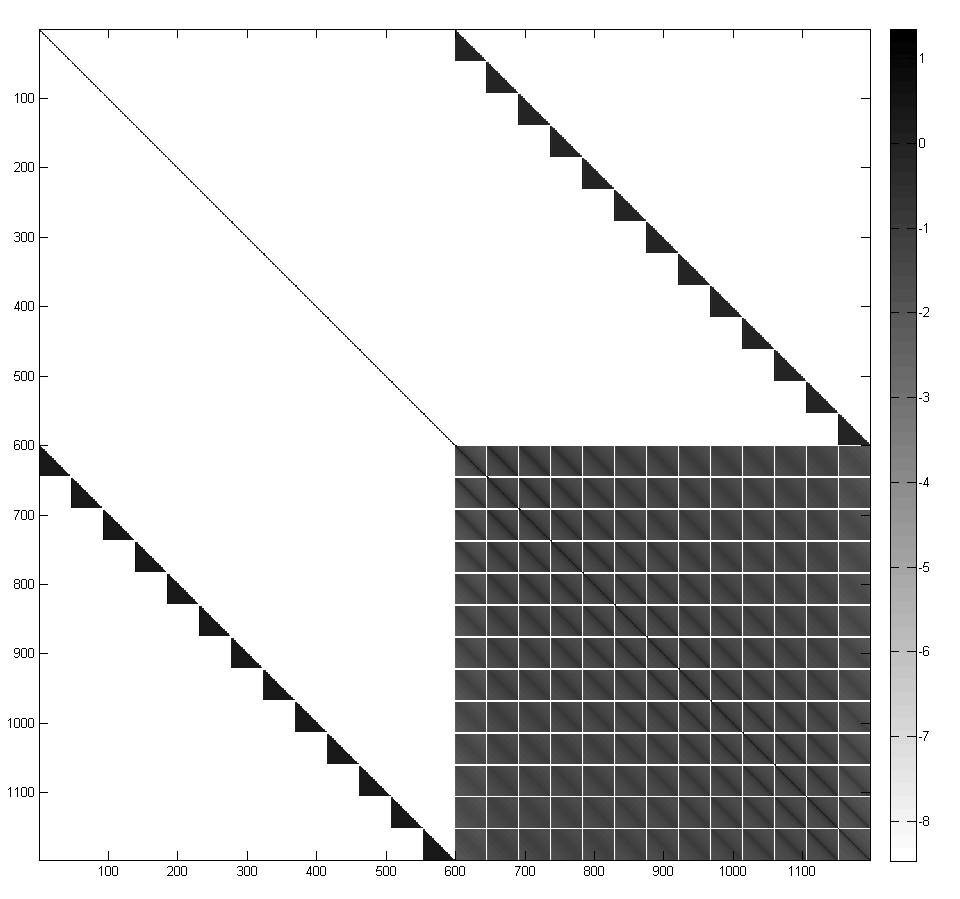}}\\
\subfloat[Lower-right submatrix of the nonlinear Jacobian]{\label{fig:JacVisBR}\includegraphics[width=.45\linewidth]{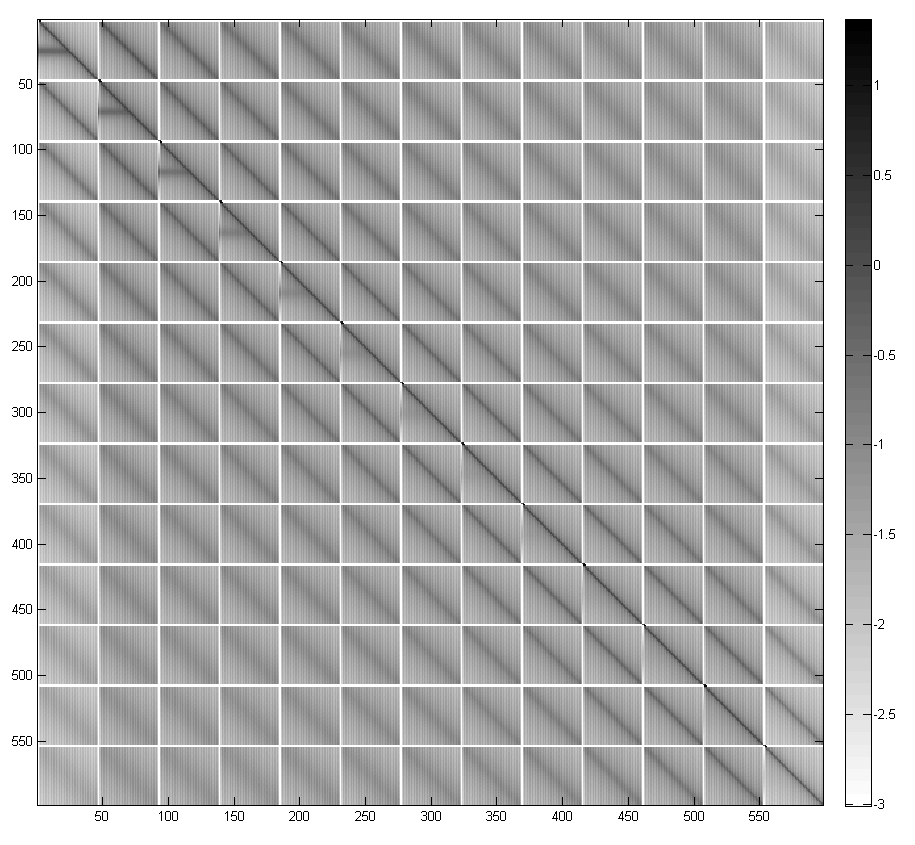}}
\subfloat[Lower-right submatrix of the linear Jacobian]{\label{fig:JacLinVisBR}\includegraphics[width=.45\linewidth]{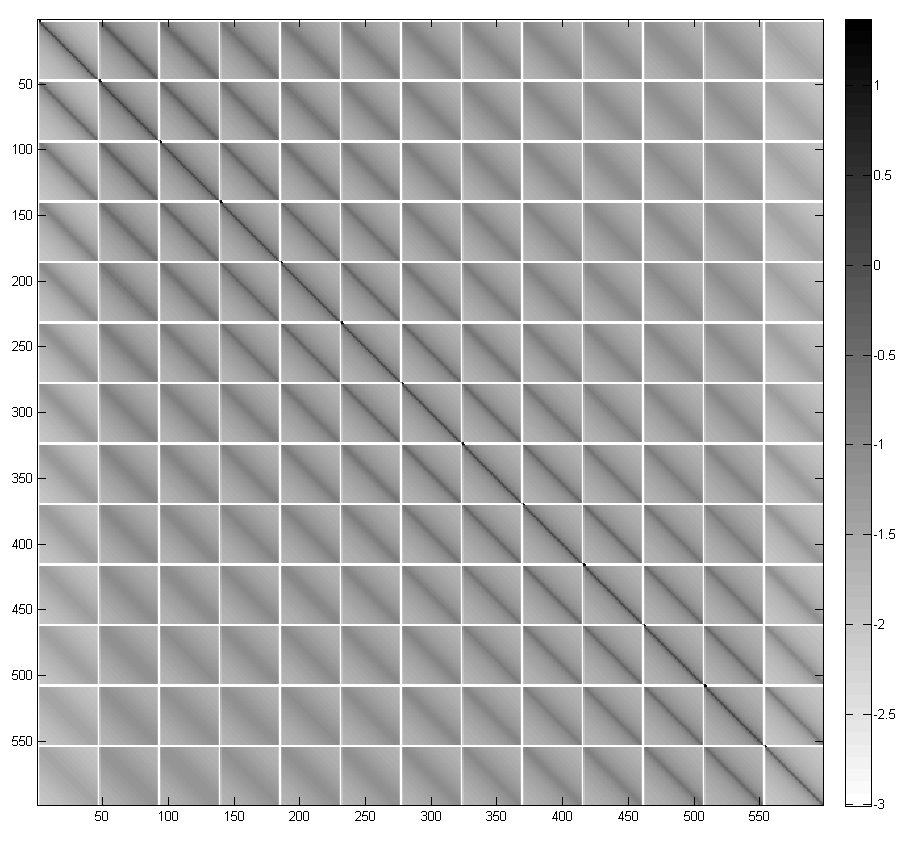}}
\caption[Visualisation of Jacobians]{A visualisation of the magnitude of the (a) nonlinear and (b) linear Jacobian entries.  A close-up of the lower-right submatrices are shown in (c) and (d) , respectively.  Computed for $\textbf{u}_k=\textbf{u}_0$ (which corresponds to an initial guess of a flat surface) with parameters $\epsilon=1$ and $F=0.7$, on a log scale: $\mathrm{log}_{10}|J_{i,j}|$ for all $i$, $j$. Each element is assigned a shade based on its value: the larger the value, the darker the shade.}
\label{fig:TwoJacVis}
\end{figure}
\FloatBarrier

\begin{figure}[htb]
\begin{center}
\subfloat[$\sigma(\textbf{J}_k)$]{\label{fig:JacEig}\includegraphics[width=.5\linewidth]{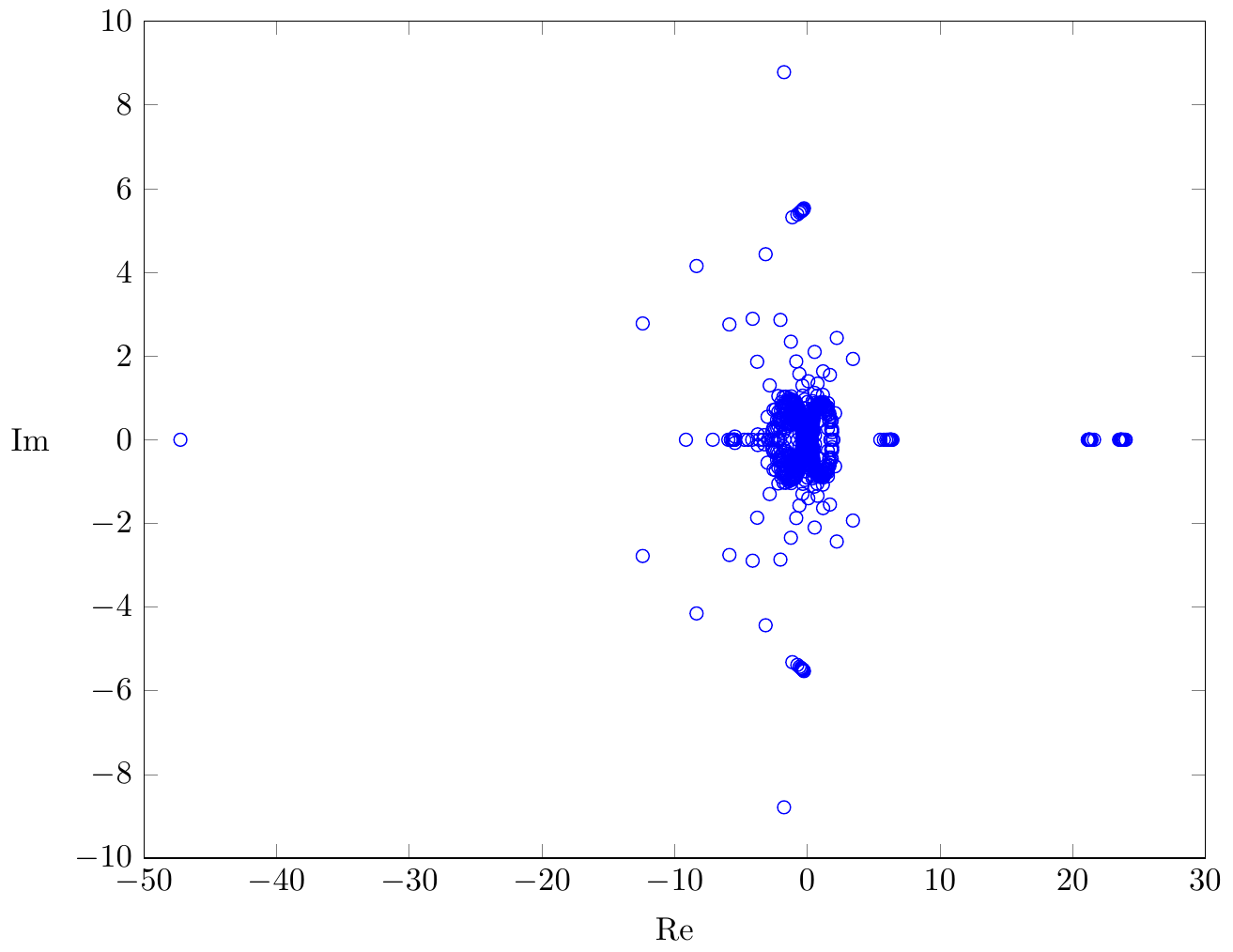}}
\subfloat[$\sigma(\textbf{J}_k\textbf{P}^{-1})$]{\label{fig:JacXinvLinEig}\includegraphics[width=.5\linewidth]{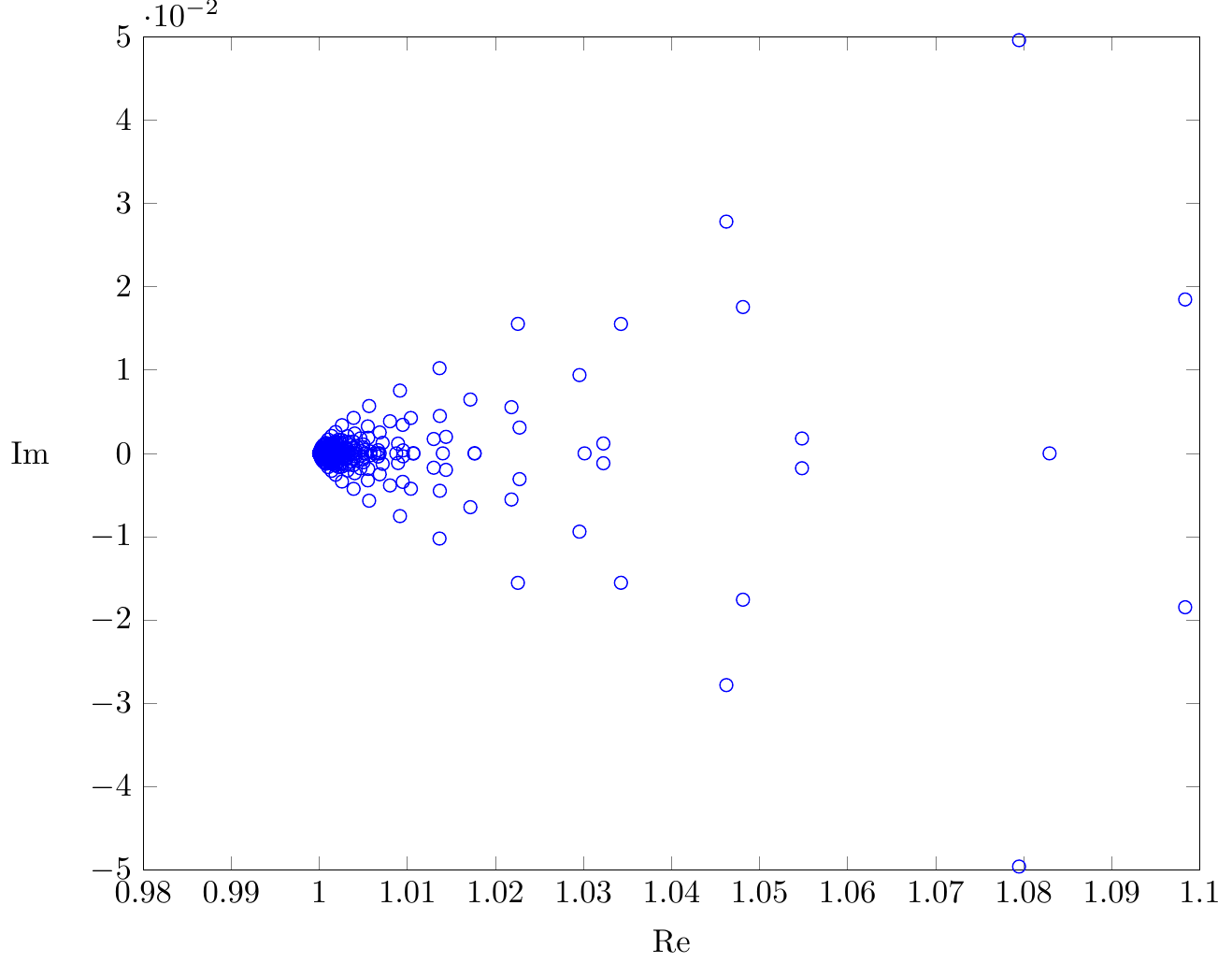}}
\caption{A plot of the spectrum for $\textbf{J}_k$ and $\textbf{J}_k\textbf{P}^{-1}$ on a $45\times 13$ mesh with $\Delta x=0.4$, $\Delta y=0.4$, $x_0=-9$ with $\textbf{u}_k=\textbf{u}_0$ and non-dimensional parameters $\epsilon=1$ and $F=0.7$. Here, $\textbf{P}$ is the full linear preconditioner (which is dense in the lower-right submatrix).}
\label{fig:TwoJacEig}
\end{center}
\end{figure}

In Figure~\ref{fig:TwoJacVis}, the Jacobian matrix for the full nonlinear problem ((a) ``nonlinear Jacobian'') for $\textbf{u}_k={\bf u}_0$ with parameters $\epsilon=1$ and $F=0.7$ is compared to its counterpart for the linear problem ((b) ``linear Jacobian'') by means of the magnitude of their entries.  The comparison confirms that, although there are slight differences in the magnitude of these entries (in particular, the grey triangular regions near the diagonal in the upper-left submatrix in Figure~\ref{fig:TwoJacVis}(a) do not appear in Figure~\ref{fig:TwoJacVis}(b)), the general structure of the two matrices is the same.  The eigenvalue spectra of the nonlinear Jacobian before and after preconditioning with the linear Jacobian are exhibited in Figure~\ref{fig:TwoJacEig}.  The figure reveals that the application of the preconditioner has resulted in a tight clustering of the eigenvalues around unity, confirming its effectiveness.

While the linear Jacobian is significantly cheaper to compute than its nonlinear counterpart, its lower-right submatrix is nonetheless fully dense, which would ultimately limit the number of mesh nodes that could be used in the discretisation due to storage and factorisation considerations.  Therefore, we focus attention on the lower-right submatrix of the two Jacobians (Figure~\ref{fig:TwoJacVis} (c), (d)), which reveals that the magnitudes of the entries decay with distance from the main block diagonal. This observation suggests using a block-banded approximation to this portion of the matrix for our preconditioner, whereby we keep only the nonzero entries of the lower-right submatrix of the linear Jacobian within a stated block bandwidth $b$, with block sizes $(N+1)\times(N+1)$. By varying this bandwidth, the sparsity of the preconditioner can be controlled such that the storage and factorisation costs are manageable. The method of storing, factorising and applying the preconditioner is outlined in \ref{appendixC}.

In Figure \ref{fig:ThreeBBandEigenPlots} we illustrate that even with block bandwidth $b = 1$ (that is, a block diagonal approximation), the linear Jacobian still functions effectively as a preconditioner, providing the required eigenvalue clustering.  The tightness of this clustering can be further improved by increasing the bandwidth, as the results for $b = 3$ and $b = 5$ confirm.

\begin{figure}[htb]
\centering
\subfloat[nonzero elements for $b=1$]{\label{fig:JacXinvBBand0Spy}\includegraphics[width=.3\linewidth]{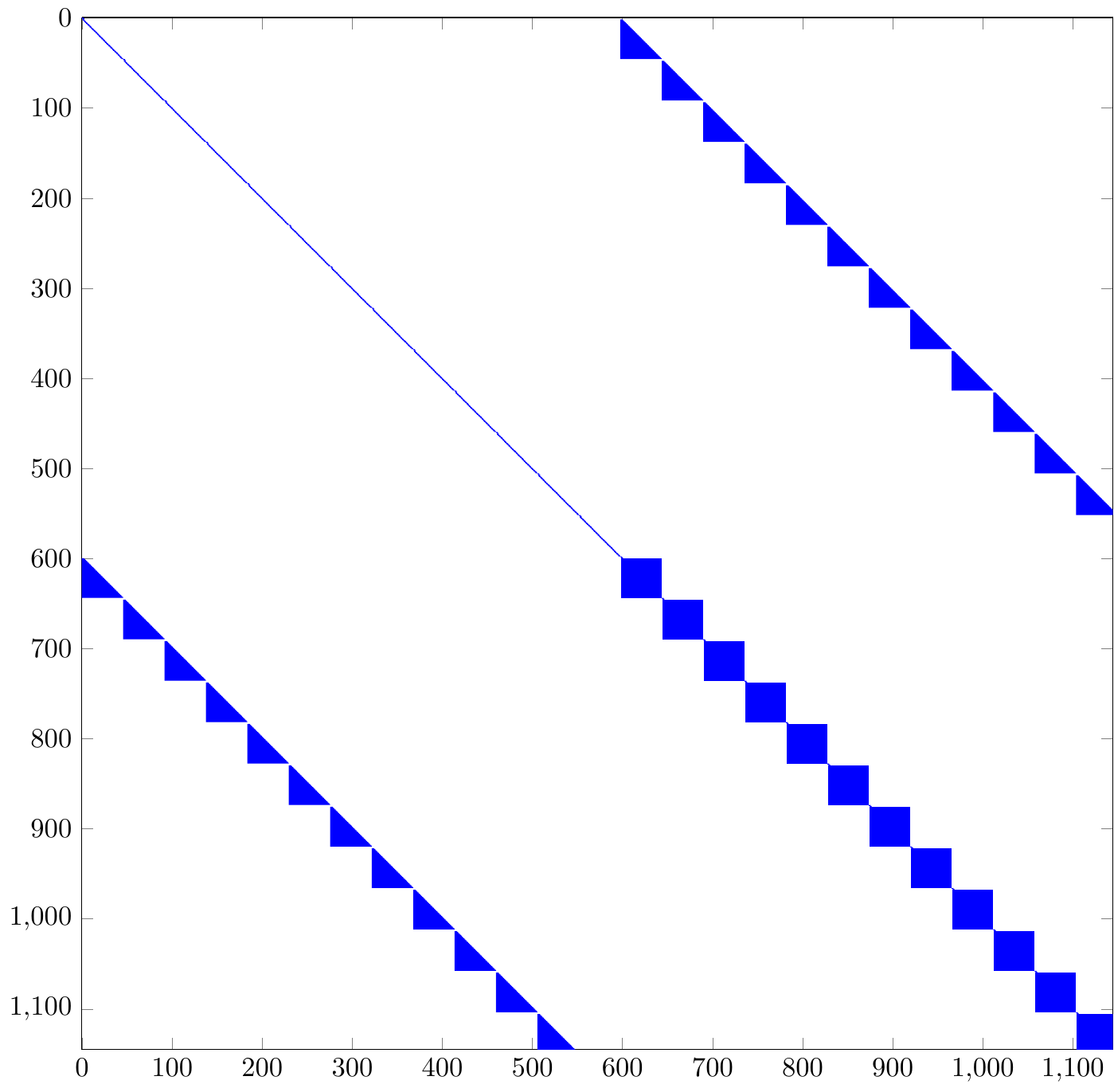}}
\hfil
\subfloat[nonzero elements for $b=3$]{\label{fig:JacXinvBBand1Spy}\includegraphics[width=.3\linewidth]{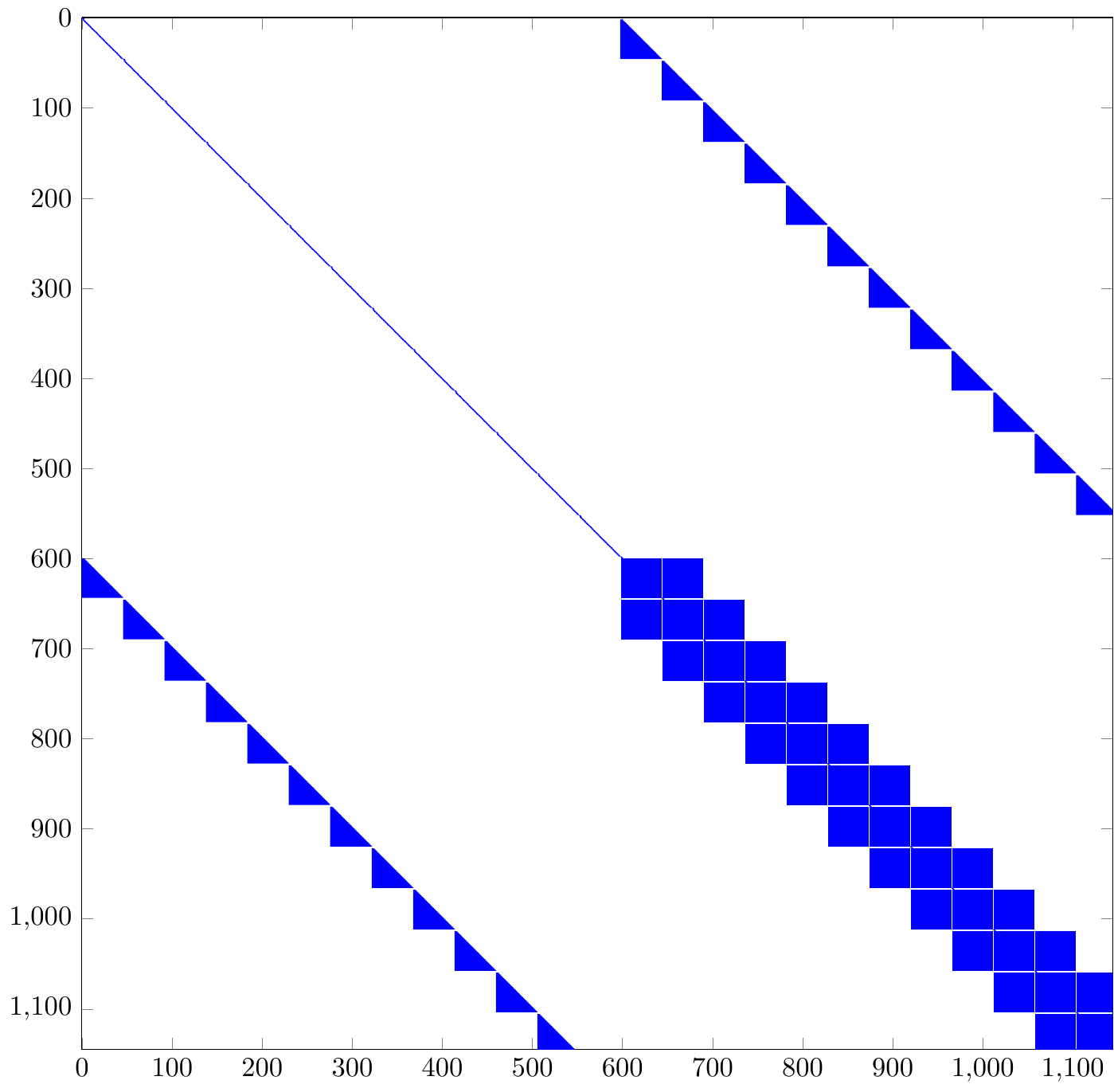}}
\hfil
\subfloat[nonzero elements for $b=5$]{\label{fig:JacXinvBBand2Spy}\includegraphics[width=.3\linewidth]{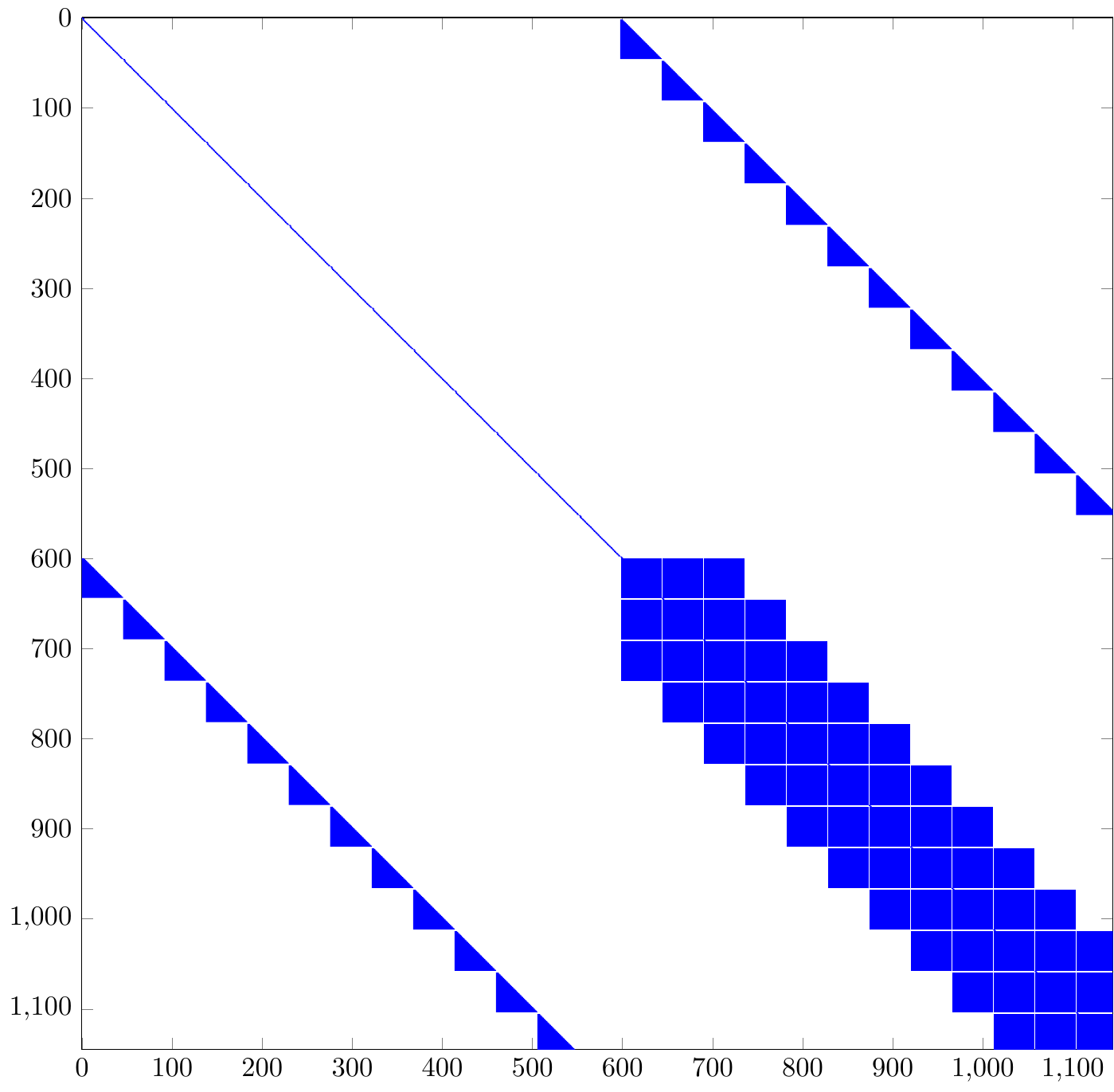}}\\
\subfloat[spectrum $\sigma(\textbf{J}_k\textbf{P}^{-1})$ for $b=1$]{\label{fig:JacXinvBBand0Eig}\includegraphics[width=.3\linewidth]{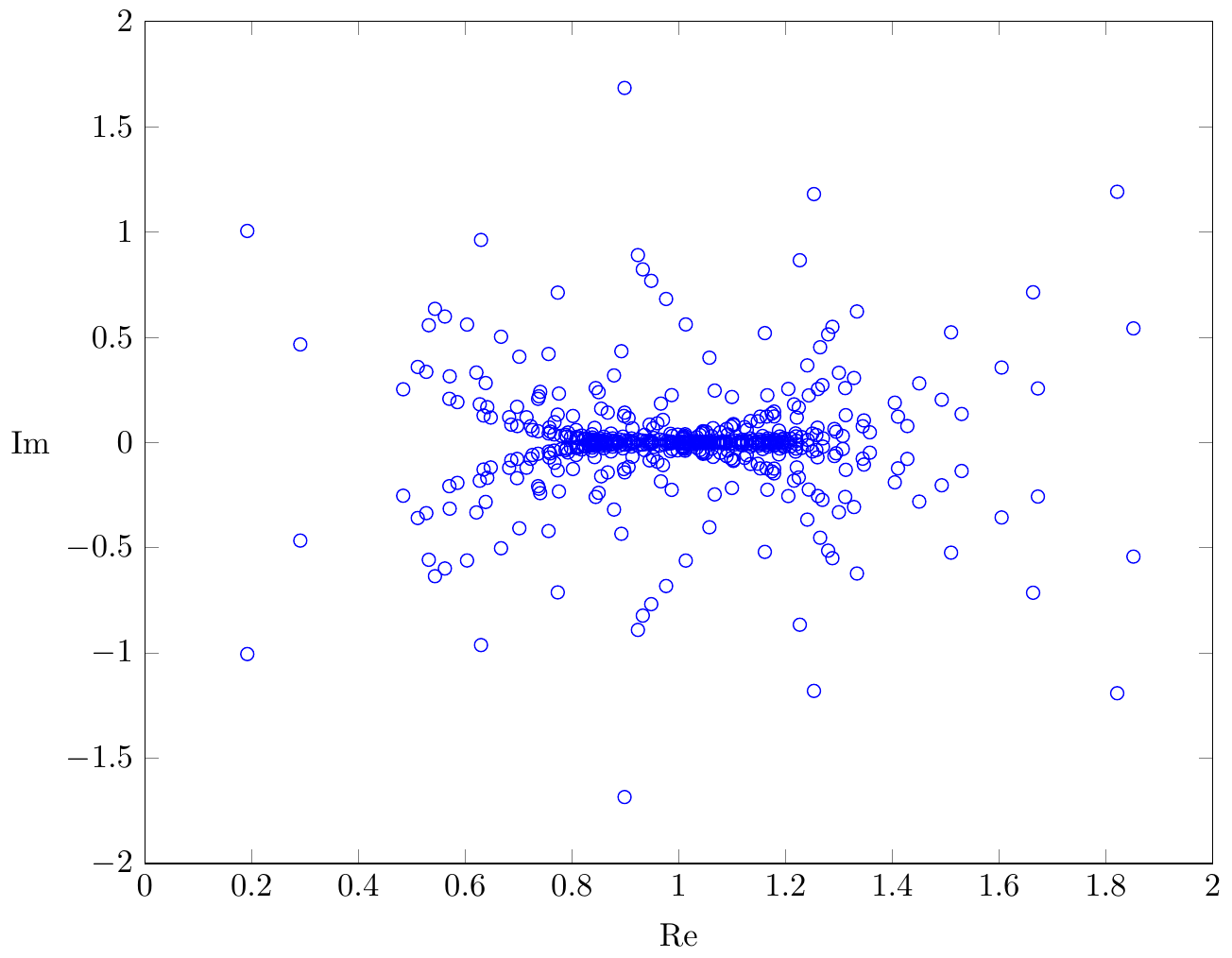}}
\hfil
\subfloat[spectrum $\sigma(\textbf{J}_k\textbf{P}^{-1})$ for $b=3$]{\label{fig:JacXinvBBand1Eig}\includegraphics[width=.3\linewidth]{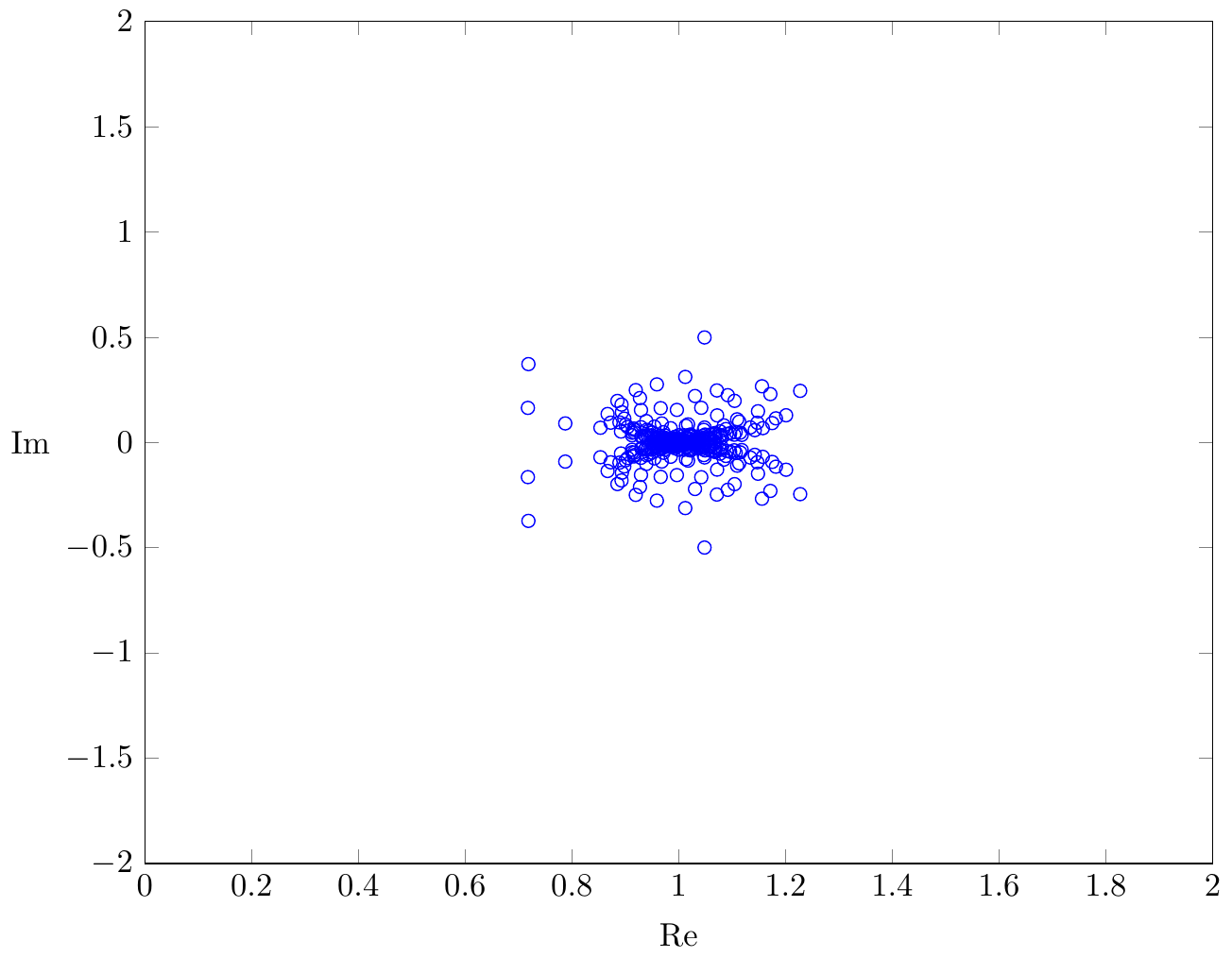}}
\hfil
\subfloat[spectrum $\sigma(\textbf{J}_k\textbf{P}^{-1})$ for $b=5$]{\label{fig:JacXinvBBand2Eig}\includegraphics[width=.305\linewidth]{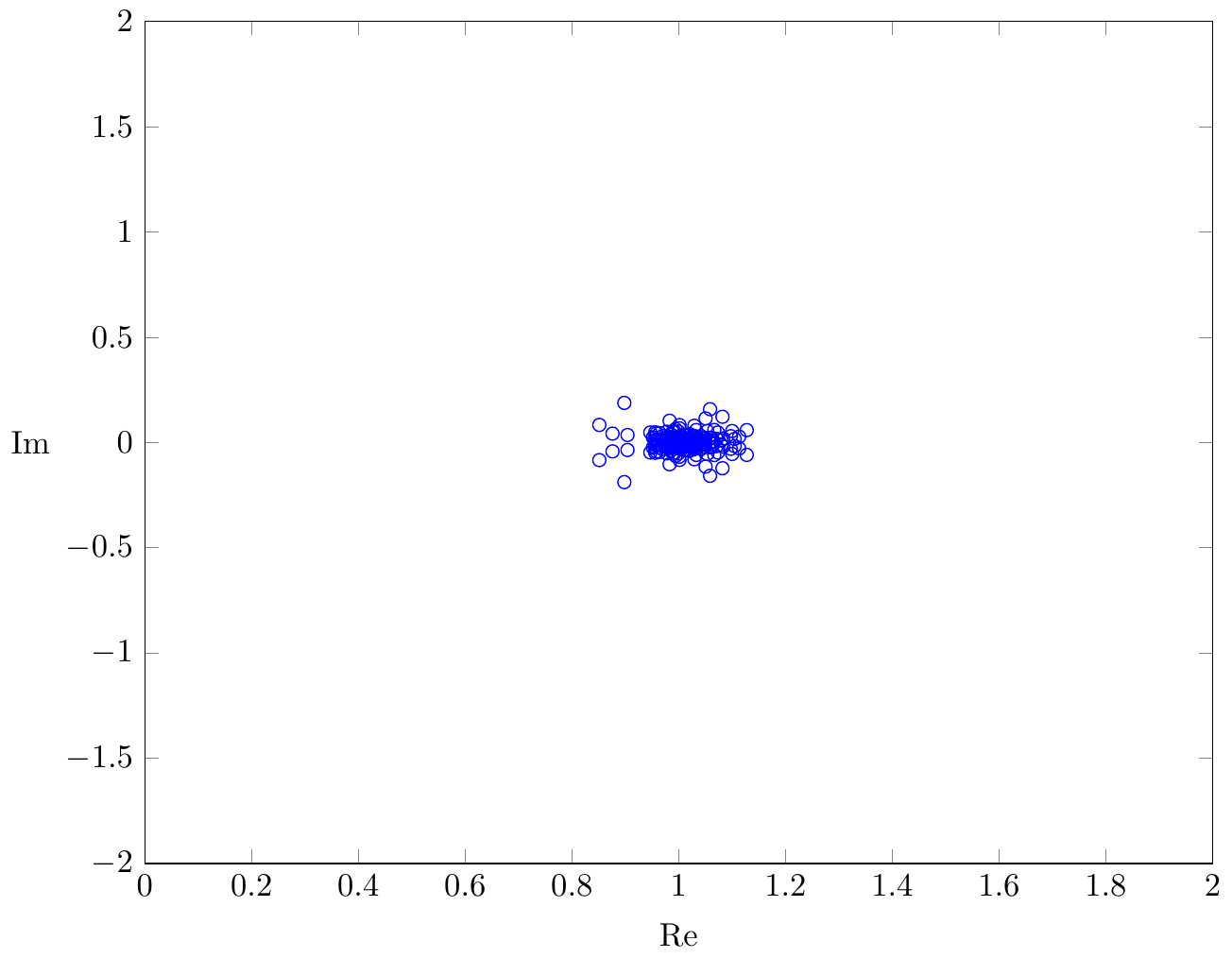}}\\
\caption{Location of nonzero elements in the preconditioner $\textbf{P}$ for bandwidths (a) $b=1$, (b) $b=3$, and (c) $b=5$, all computed for a $45\times 13$ mesh with $\Delta x=0.4$, $\Delta y=0.4$, $x_0=-9$ and non-dimensional parameters $\epsilon=1$ and $F=0.7$.  Associated plots of the spectrum of $\textbf{J}_k\textbf{P}^{-1}$ for $\textbf{u}_k=\textbf{u}_0$: (d) $b=1$, (e) $b=3$, (f) $b=5$.}
\label{fig:ThreeBBandEigenPlots}
\end{figure}
\section{Results}
\label{sec:results}
We have computed solutions using both a standard desktop computer\footnote{Intel Core i7-2600 CPU with 3.40 GHz processor and 8 GB of system memory} with all code written in MATLAB, and using a more powerful workstation with GPU accelerator\footnote{2x Intel Xeon E5-2670 CPUs with 2.66 GHz processor, M2090 Nvidia Tesla GPU and 124 GB of system memory} using a mixture of MATLAB and CUDA (Compute Unified Device Architecture) code.
In all cases the KINSOL \cite{hindmarsh05} implementation of the Jacobian-Free Newton-Krylov method was used.  In the following, recall that an $N\times M$ mesh involves $N$ grid points in the $x$ direction and $M$ grid points in the $y$ direction.

\subsection{Desktop Computer}
We present results obtained by solving our system of nonlinear equations on a typical desktop computer for a contemporary mesh ($91\times 31$, $\Delta x=\Delta y=0.3$, $x_0=-9$) as well as for a significantly finer mesh ($361\times 121$, $\Delta x=\Delta y=0.15$, $x_0=-14$).  The parameter values we focus on are $F=0.7$ and $\epsilon=1$, which are representative of a moderately small Froude number and a moderately nonlinear flow regime.

\begin{figure}[htb]
\centering
\subfloat[Orthographic view]{\label{fig:4a}\includegraphics[width=.5\linewidth]{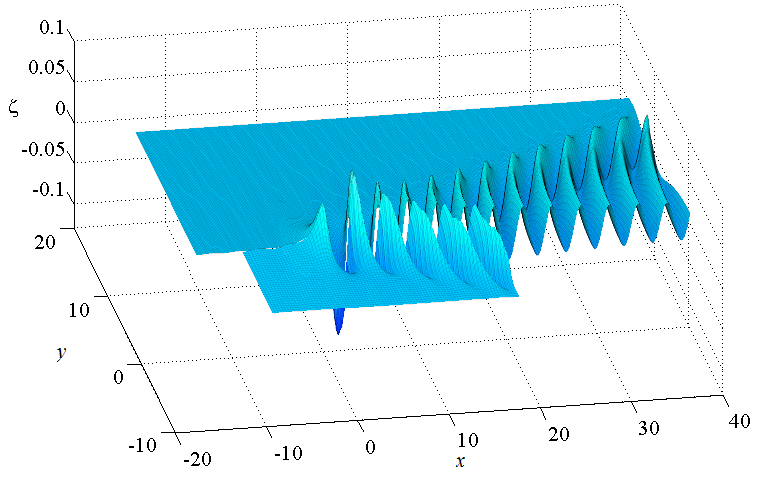}}
\subfloat[Plan view]{\label{fig:4b}\includegraphics[width=.5\linewidth]{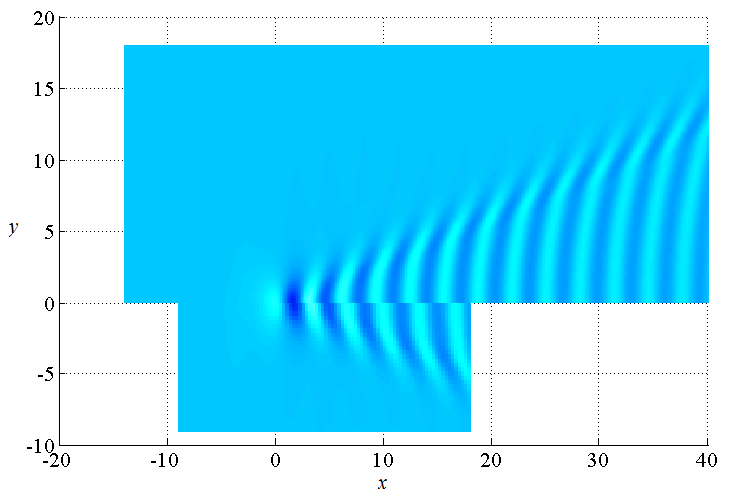}}
\caption{A comparison of free surface profiles for $F=0.7$ and $\epsilon=1$
computed on two different grids in (a) orthographic view and (b) plan view. The surface on the near side (bottom) corresponds to a $91\times 31$ mesh with $\Delta x=0.3$, $\Delta y=0.3$, $x_0=-9$, while the surface on the far side (top) is for a $361\times 121$ mesh with $\Delta x=0.15$, $\Delta y=0.15$, $x_0=-14$.  Both solutions are computed on a desktop PC.}
\label{fig:4}
\end{figure}

For the contemporary mesh, the resulting problem size is sufficiently small that the full preconditioner (without taking the banded approximation) can be formed and factorised without difficulty on today's desktop machines.  Using the Jacobian-free Newton-Krylov method with this dense preconditioner, the solution was obtained in under 26 seconds.  Calculating numerical solutions like this one in such a small time is useful for exploring the effect of different parameter values on the free surface; however, as can be seen in Figure~\ref{fig:4}, the resulting surface is rather coarse, and does not reveal much detail of the wave pattern.

By using the block-banded preconditioner with our Jacobian-free Newton-Krylov method, we are able to compute the solution on the much finer mesh ($361\times 121$) in under 75 minutes on the desktop computer.  A block bandwidth of $b = 31$ is used for the Jacobian, which means it essentially fills all of the available system memory.  This level of mesh refinement represents a comfortable size of problem for the given machine, and produces a free surface profile that is significantly smoother than the one computed with a $91\times 31$ mesh (again, see Figure~\ref{fig:4}).  With a modest degree of further refinement, the problem may still be solved on the desktop computer, however the effectiveness of the preconditioner is reduced owing to the limited number of bands that can be accommodated in memory.

\subsection{Workstation with GPU Accelerator}
\begin{figure}[htb]
\centering
\subfloat[A close up of the wave pattern.]{\label{fig:721x241_ep1_Fr07_closeUp} \includegraphics[width=.8\linewidth]{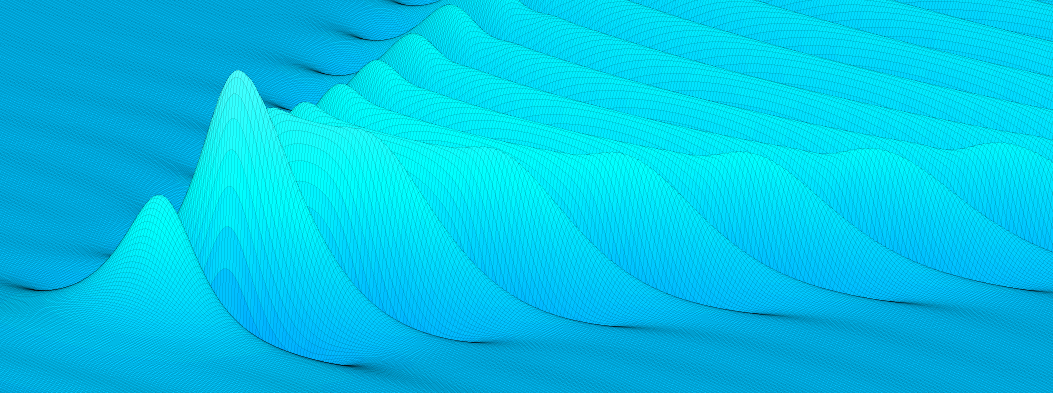}}\\
\subfloat[The free surface profile over the full truncated domain.]{\label{fig:721x241_ep1_Fr07_full} \includegraphics[width=.52\linewidth]{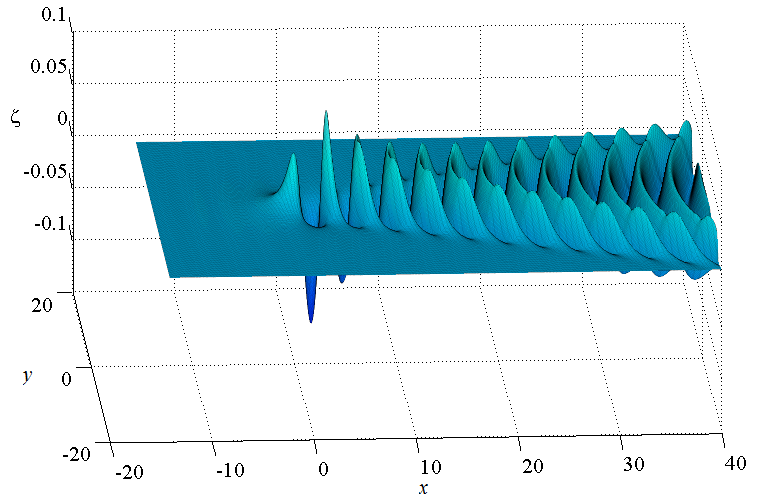}}
\subfloat[A contour plot, where contours are shown only for positive wave heights to avoid confusion.]{\label{fig:721x241_ep1_Fr07_contour}\includegraphics[width=.46\linewidth]{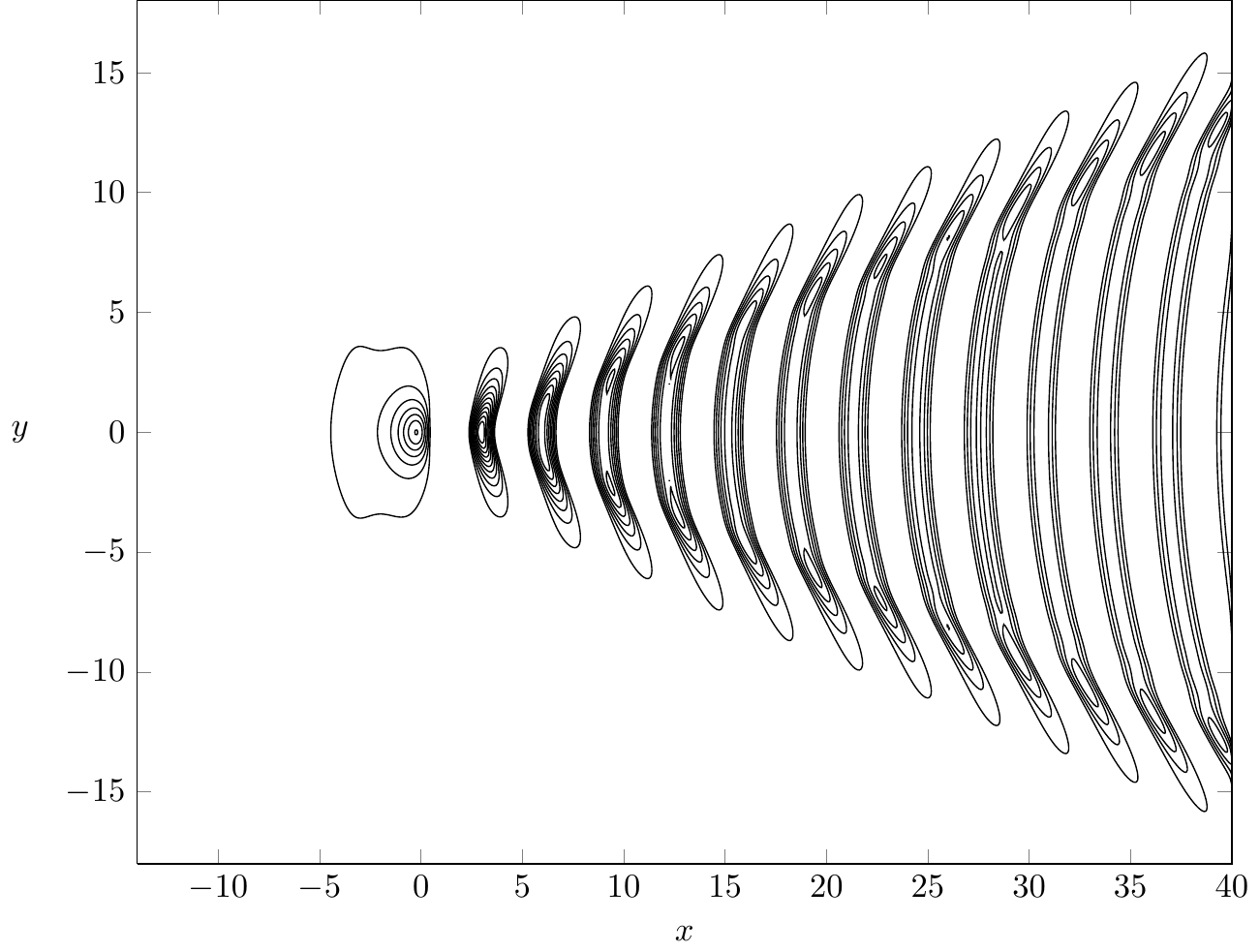}}
\caption{Perspectives of the free surface for $F=0.7$ and $\epsilon=1$,
computed on a $721\times 241$ mesh with $\Delta x=0.075$, $\Delta y=0.075$, $x_0=-14$. This solution was computed on a workstation with GPU acceleration.}
\label{fig:5}
\end{figure}

By coding the nonlinear discretisation in CUDA and executing each evaluation (hereafter a ``function evaluation'') on the GPU, we were able to significantly accelerate the computations as demonstrated in Table~\ref{tab:FuncEvalTimes}.  Here we are experiencing an approximately 25 times speed up in function evaluation times over the multicore MATLAB code for the larger meshes. This leads to a reduced overall runtime,  for example, calculating the solution on the same $361\times 121$ mesh with GPU acceleration took only 3.5 minutes. This dramatic reduction in computational time coupled with the extra system memory available on the workstation allowed us to produce solutions on much finer meshes in a practical amount of time. Our most detailed solution using a $721\times 241$ mesh with $\Delta x=\Delta y=0.075$ and $x_0=-14$, was computed in 1.5 hours.  The corresponding free surface profile is illustrated in Figure~\ref{fig:5}.
\begin{table}[htb]
\begin{center}
\begin{tabular}{c|c|c|c}
Mesh & Multicore MATLAB & Multicore MATLAB & MATLAB with  \\
& on desktop& on workstation& GPU on workstation\\
\hline $91\times 31$ &  1.43 &  0.87 &  0.02 \\
$181\times 61$ &   11.26 &  4.38 &  0.18 \\
$271\times 91$ &   54.47 &  20.40 &  0.83 \\
$361\times 121$ &  169.46 &  64.35 &  2.52 \\
$451\times 151$ &  410.34 &  153.91 &  6.04 \\
\end{tabular}
\end{center}
\caption{A comparison of the function evaluation times using multicore MATLAB on the desktop PC and the workstation with and without GPU acceleration for different meshes with parameters $\epsilon=1$ and $F=0.7$. Time is in seconds.}
\label{tab:FuncEvalTimes}
\end{table}

\subsection{Towards grid independence}

As mentioned in the Introduction, a common procedure in the free-surface literature is to explore grid independence by computing solutions on a given truncated domain with more grid points (twice as many, say) and visually comparing the free surface profiles to test whether the grid refinement has not significantly altered the solution. Similarly, authors often keep the spatial increment the same and increase the size of the truncated domain (make it twice as long, say), again to test whether the solution changes. For steady two-dimensional flows, this exercise is reasonably straight forward (in principle), as the free surface profile is a curve.  Examples of these tests for two-dimensional problems that involve a downstream wavetrain can be found in \cite{mccue99,mekias91,zhang96}, all of which were published at a time when demonstrating grid independence was still a difficult issue.

More recently, equivalent tests of grid independence have been attempted for three-dimensional flows past disturbances \cite{parau02,parau07b,parau07c}.  In this case, as the wave pattern is a two-dimensional surface, the domain was divided in half, with one part showing a solution computed with a particular grid, and the other part with a solution computed with a more refined or extended grid.  Such a comparison is also given in Figure~\ref{fig:4}.  What we can see from Figure~\ref{fig:4} is that the solution computed on the  $91\times 31$ mesh is clearly not grid independent, as the more refined surface corresponding to a  $361\times 121$ mesh appears to be different, even on this larger scale.  We have conducted the same comparison exercise for a variety of parameter sets and meshes for our problem, and conclude that the number of grid points used presently in the literature (for a range of very similar problems) is not nearly enough for authors to claim their solutions are grid independent.  Similarly, noting that P\u{a}r\u{a}u and coauthors \cite{parau02,parau07c}  call these visual comparisons `accuracy checks', we would not say that solutions computed with contemporary meshes are accurate.  Of course it is understandable that these coarse meshes have been used in published studies, given the dense nature of the nonlinear Jacobian, the lack of a Jacobian-free approach such as we are using here, and computational power.  We hope that the algorithms presented here will allow much more accurate computations in the future.

\begin{figure}[htb]
\begin{center}
\includegraphics[width=0.85\linewidth]{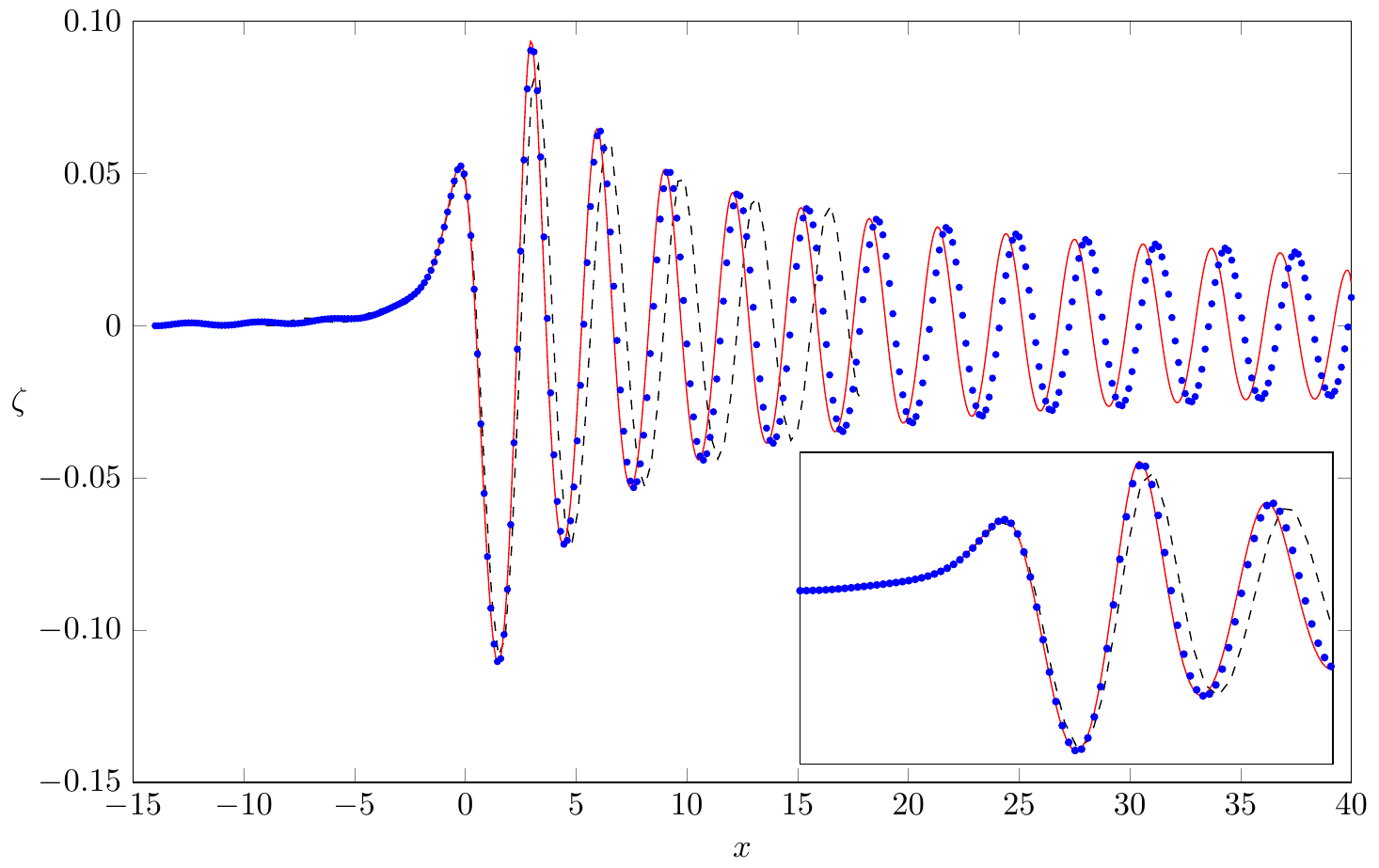}
\caption{A comparison of the centreline profiles for $F=0.7$ and $\epsilon=1$ computed on three different grids. The dashed curve has 91 nodes in the $x$-direction with $\Delta x=0.3$.  The surface made up by solid circles has 361 nodes with $\Delta x=0.15$.  Note that each circle here represents an actual grid point (the illusion of uneven grid spacing is due to the vastly different scales in the $x$ and $z$ directions).  The solid curve has 721 nodes with $\Delta x=0.075$. The inset shows a close up of this comparison near $x=0$.}
\label{fig:CentrelineEp1Fr7All}
\end{center}
\end{figure}

Another obvious approach for observing the degree of grid independence is to plot the centreline of the free surface ($z=\zeta(x,0)$) for a number of difference meshes, as shown in Figure \ref{fig:CentrelineEp1Fr7All}.  In addition to the $91\times 31$ and $361\times 121$ meshes used in Figure~\ref{fig:4}, we have also included the centreline plot for the $721\times 241$ mesh used in Figure~\ref{fig:5}.  Recall that this latter mesh was implemented a workstation with GPU acceleration.  We see there is quite good agreement between the solutions for the $361\times 121$ and $721\times 241$ meshes, at least over the first four or five wavelengths.  Further downstream the amplitudes of the waves appear to agree well, but the actual wavelength is slightly out.  This comparison suggests that while we can not yet claim our solutions will not be affected by further grid refinement, we argue that meshes of the order of $361\times 121$ and $721\times 241$ are required for solutions to begin to appear independent of the mesh spacing and truncation.

It is worth making some comments about the truncation errors we discussed at the end of Section~\ref{sec:numerical}.  First, we note that truncating the domain upstream at $x=x_1$ has the effect of introducing very small spurious (almost two-dimensional) waves throughout the domain.  These may be seen in Figures~\ref{fig:4} and \ref{fig:5}, both ahead of the source and also outside of the Kelvin wedge.  This numerical artefact has been an issue for two-dimensional flows for many years, and the associated spurious waves have been eliminated by employing a variety of upstream boundary conditions \cite{parau02,zhang96}.  A detailed discussion for two-dimensional flows is given by Grandison \& Vanden-Broeck~\cite{grandison06}.  In our scheme, the enforcement of the radiation condition via (\ref{eqn:numUpRadiation}) has the effect of dramatically reducing the size of these spurious waves (the coefficient $n$ is chosen based on these observations).  This issue deserves further attention.

Further, we note any truncation of the domain at $x=x_N$ will introduce errors in the system, as the contribution from the wavetrain to the integrals for $x>x_N$ will be ignored.  Visually, we can see in Figure~\ref{fig:CentrelineEp1Fr7All} that the final wavelength of the free surface seems affected by this truncation.  Again, strategies have been developed to deal with these errors in much simpler two-dimensional problems \cite{grandison06}, and similar work is needed for the types of three-dimensional flows considered here.

\subsection{Details of wave patterns}

The free-surface profiles presented in Figures \ref{fig:4}-\ref{fig:5} are computed for the moderately small value of the Froude number, $F=0.7$. In this regime, the transverse waves, which run perpendicular to the flow direction, are prominent. These are the waves we observe in the centreline plot in Figure~\ref{fig:CentrelineEp1Fr7All}. The other type of waves are the divergent waves, whose crests appear to form ridges pointing diagonally away from the source. The amplitude of the transverse waves decays as $x$ increases, leaving the divergent waves to dominate at larger distances away from the source. It is the divergent wave pattern that characterises the well-known V-shaped Kelvin wake.

\begin{figure}[htb]
\centering
\subfloat[A close up of the wave pattern.]{\label{fig:721x241_ep1_Fr14_closeUp} \includegraphics[width=.8\linewidth]{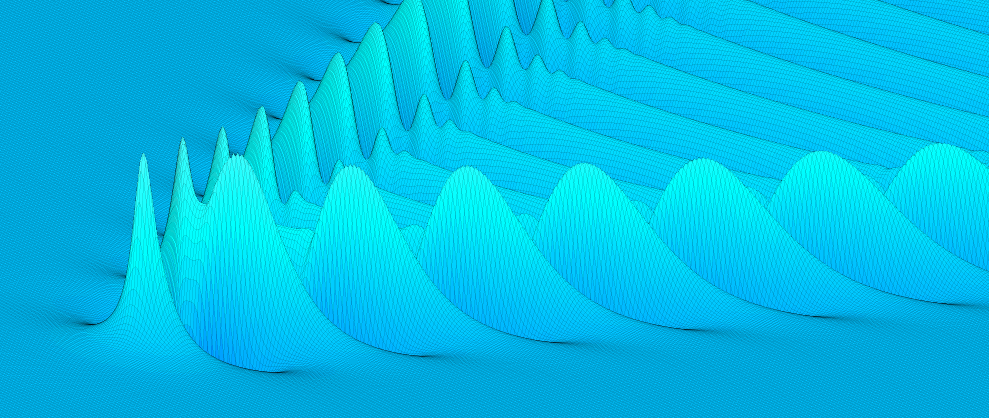}}\\
\subfloat[The free surface profile over the full truncated domain.]{\label{fig:721x241_ep1_Fr14_full} \includegraphics[width=.52\linewidth]{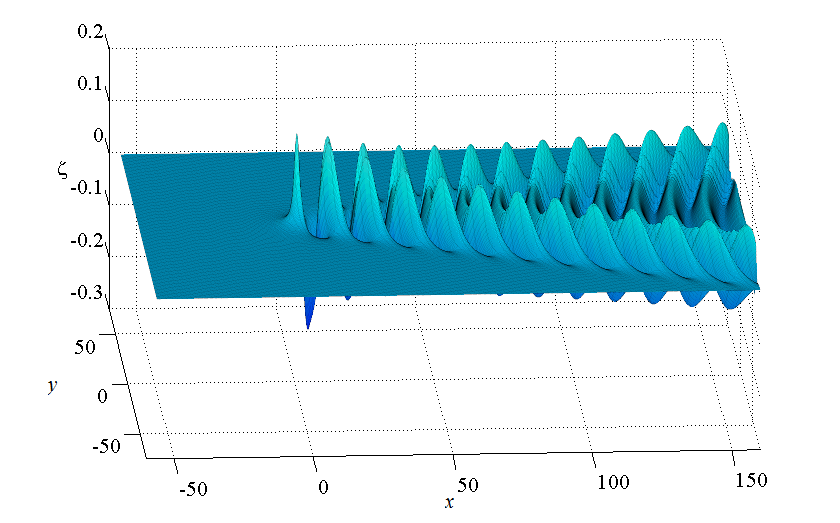}}
\subfloat[A contour plot, where contours are shown only for positive wave heights to avoid confusion.]{\label{fig:721x241_ep1_Fr14_contour}\includegraphics[width=.46\linewidth]{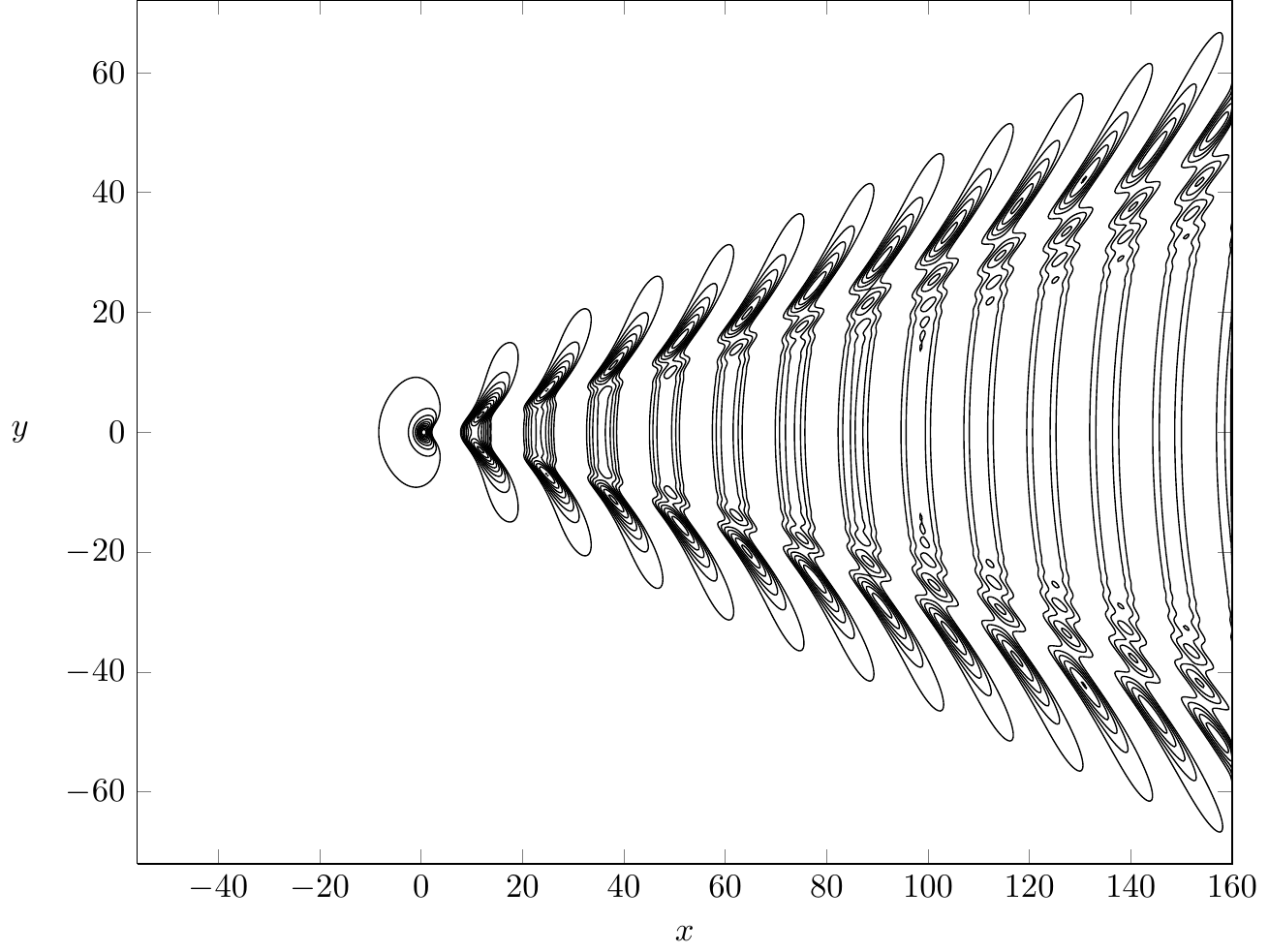}}
\caption{Perspectives of the free surface for $F=1.4$ and $\epsilon=1$,
computed on a $721\times 241$ mesh with $\Delta x=0.3$, $\Delta y=0.3$, $x_0=-56$. This solution was computed on a workstation with GPU acceleration.}
\label{fig:7}
\end{figure}

A free-surface profile computed for $F=1.4$ and $\epsilon=1$ is presented in Figure~\ref{fig:7}. For this moderately large Froude number, we see that the divergent waves dominate closer to the source, making it more difficult to view the transverse waves. Note that the wavelength of the transverse waves increases with Froude number, which means we need to truncate further downstream for larger Froude numbers in order to capture the same amount of detail. The solution in this figure was computed using a mesh of $721\times 241$ on a workstation with GPU acceleration. With this resolution, we can see fine details of the surface in part (a) of the Figure.

\section{Discussion}
\label{sec:discussion}

We have considered the fully nonlinear problem of the free-surface flow past a submerged point source.  Following Forbes~\cite{forbes89}, we apply a boundary-integral technique based on Green's second formula to derive a singular integro-differential equation for the velocity potential $\Phi(x,y,z)$ and the shape of the surface $z=\zeta(x,y)$.  This equation, together with Bernoulli's equation, is discretised and satisfied at midpoints on a two-dimensional mesh.  The resulting system of nonlinear algebraic equations is solved using Newton's method.  In the past, numerical approaches of this sort were hindered by the fact that the Jacobian matrix in Newton's method is dense.  Our contribution is to apply a Jacobian-free Newton-Krylov method to solve the nonlinear system, thus avoiding the need to ever form or factorise the Jacobian.  As such, we are able to use much finer meshes than used in the past by other authors.  Further, in order to ensure efficiency, we use a banded matrix preconditioner whose nonzero entries come from the linearised problem.  Finally, we code the function to run efficiently on a GPU, to greatly speed up function evaluation times.  The resolution of the mesh we use is now essentially up to the standard of many two-dimensional schemes published in the literature.

As discussed in the Introduction, the problem of flow past a source singularity can be thought of as a building block for more complicated configurations such as flow due to a steadily moving applied pressure distribution (like a hovercraft), a thin ship hull, or a submerged body (like a submarine).  The next stage in this research is to adapt the present techniques for these more complicated flows.  We expect that the key ideas developed in this paper will generalise in a straightforward manner, provided there is a natural linearised version of the problem at hand.  With the accuracy and efficiency of our approach, one may be able to devise appropriate optimisation schemes for designing ship hulls with minimal resistance, and so on. Our approach should also translate to time-dependent problems, such as the study by P\u{a}r\u{a}u et al.~\cite{parau10}, who apply a similar boundary integral approach, discretised with $60\times 40$ meshes, to solve for time-dependent flows past a pressure distribution (see \cite{dias06,fochesato06,grilli01} for a thorough discussion of further issues that arise in time-dependent problems).  We leave all this work for further study.

With the degree of accuracy our numerical schemes allow, we are now in a position to explore the effect of strong nonlinearity on the wave pattern, as has been done extensively in the two-dimensional analogue.  For example, as the nonlinearity in a steady ship wave problem increases (for our problem this tendency comes from increasing $\epsilon$), the waves will become more nonlinear in shape, perhaps with sharper crests.  Given the flow is steady, we expect that the waves will ultimately ``break'' when the most nonlinear wave reaches a limiting configuration (this occurs when the highest wave crest reaches the dimensionless height $F^2/2$). While this general behaviour is well understood for two-dimensional waves, with studies of highly nonlinear waves producing highly accurate calculations of near-breaking waves \cite{cokelet77,dallaston10,Lukomsky02,schwartz74,williams81} (the breaking point corresponding to the Stokes limiting configuration with a $120^\circ$ angle at the wave crest), the highly nonlinear regime for fully three-dimensional problems is relatively unexplored.  Indeed, the extra dimension makes the pattern structure much more complicated, and so it is not always obvious what part of the domain will break first.  As such, the challenge of generalising the two-dimensional results to three dimensions remains.

\section*{Acknowledgement}
\noindent SWM acknowledges the support of the Australian Research Council via the Discovery Project DP140100933.  The authors thank Prof.\ Kevin Burrage for the use of high performance computing facilities and acknowledge further computational resources and support provided by the High Performance Computing and Research Support (HPC) group at Queensland University of Technology.

\section*{References}
\bibliographystyle{plain}


\appendix

\section{Ordering the equations}\label{appendixA}
The left-hand side of $\textbf{E}(\textbf{u})=0$ is a vector valued function made up of six different functions taken from the numerical scheme. The free surface condition (\ref{eqn:freeSurfCond}) and boundary integral equation (\ref{eqn:IntegroEqn}) evaluated at the half mesh points $(x_{k+\frac{1}{2}},y_\ell)$ are denoted $\textbf{E}_{1_{k,\ell}}$ and $\textbf{E}_{2_{k,\ell}}$, respectively,  for $k=1,\dots,N-1$ and $\ell=1,\dots,M$. We also have the radiation conditions (\ref{eqn:numUpRadiation}) denoted:
\begin{equation*}
	\begin{aligned}
	\textbf{E}_{3_{\ell}}&=x_1((\phi_x)_{1,\ell}-1)+n(\phi_{1,\ell}-x_1),\\
	\textbf{E}_{4_{\ell}}&=x_1(\phi_{xx})_{1,\ell}+n((\phi_x)_{1,\ell}-1),\\
	\textbf{E}_{5_{\ell}}&=x_1(\zeta_x)_{1,\ell}+n\zeta_{1,\ell},\\
	\textbf{E}_{6_{\ell}}&=x_1(\zeta_{xx})_{1,\ell}+n(\zeta_x)_{1,\ell},
	\end{aligned}
\end{equation*}
for $\ell=1,\dots,M$. We order these equations as
\begin{align*}
\textbf{E}=&[\textbf{E}_{3_{1}},\textbf{E}_{4_{1}},\textbf{E}_{1_{1,1}},\dots,\textbf{E}_{1_{N-1,1}},\textbf{E}_{3_{2}},\textbf{E}_{4_{2}},\textbf{E}_{1_{1,2}},\dots,\textbf{E}_{1_{N-1,2}},\ldots,\textbf{E}_{3_{M}},\textbf{E}_{4_{M}},\textbf{E}_{1_{1,M}},\dots,\textbf{E}_{1_{N-1,M}},
\nonumber \\
&\textbf{E}_{5_{1}},\textbf{E}_{6_{1}},\textbf{E}_{2_{1,1}},\dots,\textbf{E}_{2_{N-1,1}},\textbf{E}_{5_{2}},\textbf{E}_{6_{2}},\textbf{E}_{2_{1,2}},\dots,\textbf{E}_{2_{N-1,2}},\ldots,\textbf{E}_{5_{M}},\textbf{E}_{6_{M}},\textbf{E}_{2_{1,M}},\dots,\textbf{E}_{2_{N-1,M}}]^T,
\end{align*}
which results in the Jacobian structure illustrated in Figure~\ref{fig:TwoJacVis}.

\section{The linear Jacobian}\label{appendixB}
To construct the linear Jacobian, we need to apply the same numerical discretisation outlined in Section \ref{sec:numerical} to the linear problem derived in Section~\ref{sec:linearproblem}.

The singularity in (\ref{eqn:IntegroEqnlinear}) is dealt with in the same way as with the nonlinear problem, by adding and subtracting the term (\ref{eqn:I2DashDash}), except that now $S_2(x_i,y_j;x^*_k,y^*_\ell)=K_3(x_i,y_j;x^*_k,y^*_\ell)$, which simplifies the details.  The linear system then becomes
\begin{equation}
\begin{aligned}
\textbf{E}_{1_{k,\ell}}&=\phi^*_{x_{k,\ell}}+\frac{\zeta^*_{k,\ell}}{F^2}-1,\\
\textbf{E}_{2_{k,\ell}}&=2\pi(\phi^*_{k,\ell}-x^*_k)+\frac{\epsilon}{\left({x^*_k}^2+{y^*_\ell}^2+1 \right)^\frac{1}{2}}
-\sum\limits_{i=1}^{N}\sum\limits_{j=1}^{M} w(i,j)\left[\zeta_{x_{i,j}}-\zeta^*_{x_{k,\ell}}\right]K_{3_{i,j,k,\ell}} -\zeta^*_{x_{i,j}}I,\\
\textbf{E}_{3_{\ell}}&=x_1\phi_{x_{1,\ell}}+n\phi_{1,\ell}-x_1(n+1),\\
\textbf{E}_{4_{\ell}}&=\frac{x_1}{\Delta x}\phi_{x_{2,\ell}}+(n-\frac{x_1}{\Delta x})\phi_{x_{1,\ell}}-n,\\
\textbf{E}_{5_{\ell}}&=x_1\zeta_{x_{1,\ell}}+n\zeta_{1,\ell},\\
\textbf{E}_{6_{\ell}}&=\frac{x_1}{\Delta x}\zeta_{x_{2,\ell}}+(n-\frac{x_1}{\Delta x})\zeta_{x_{1,\ell}},
\end{aligned}\label{eqn:SystemEquation}
\end{equation}

for $k=1\dots(N-1)$, $\ell=1\dots M$ where \textbf{E} is constructed from these equations and
\begin{align*}
K_{3_{i,j,k,\ell}}&=K_3(x_i,y_j;x^*_k,y^*_\ell),
\end{align*}
$I$ is given by
\begin{equation*}
I=\int\limits_{y_1}^{y_M}\int\limits_{x_1}^{x_N}
K_3\, \text{d}x\,\text{d}y,\label{eqn:IS2linear}
\end{equation*}
and $w(i,j)$ is the weighting function for numerical integration.  As before, $I$ can be evaluated exactly in terms of logarithms.

The next step is to determine how $\phi$, $\phi^*$, $\zeta$ and $\zeta^*$ depend on the unknowns in (\ref{eq:unknowns}).  We first expand the trapezoidal-rule integration of $\zeta$ in (\ref{eqn:zetaapprox}) which gives
\begin{equation*}
\zeta_{k,\ell}=\zeta_{1,\ell}+\frac{\Delta x}{2}\zeta_{x_{1,\ell}}+\Delta x\sum\limits_{i=2}^{k-1}\zeta_{x_{i,\ell}}+\frac{\Delta x}{2}\zeta_{x_{k,\ell}},\quad \text{for }k=2,\dots,N\text{, }\ell=1,\dots,M.\label{eqn:zetaApproxExpand}
\end{equation*}
Similarly, we expand $\phi$ as
\[
\phi_{k,\ell}=\phi_{1,\ell}+\frac{\Delta x}{2}\phi_{x_{1,\ell}}+\Delta x\sum\limits_{i=2}^{k-1}\phi_{x_{i,\ell}}+\frac{\Delta x}{2}\phi_{x_{k,\ell}},\quad \text{for }k=2,\dots,N\text{, }\ell=1,\dots,M.
\]
This result immediately provides the values for $\phi^*$ using two point interpolation
\[
\phi^*_{k,\ell}=\frac{1}{2}(\phi_{k,\ell}+\phi_{k+1,\ell}).
\]
Substituting this expression and its equivalent in $\zeta^*_x$ and $\zeta^*$ into (\ref{eqn:SystemEquation}) gives the resulting linear system in terms of the unknowns,
\begin{align}
\textbf{E}_{1_{k,\ell}}&=\frac{1}{2}(\phi_{x_{k,\ell}}+\phi_{x_{k+1,\ell}})+\frac{1}{F^2}\left(\zeta_{1,\ell}+\frac{\Delta x}{2}\zeta_{x_{1,\ell}}+ \Delta x\sum\limits_{i=2}^{k-1}\zeta_{x_{i,\ell}}+\frac{3\Delta x}{4}\zeta_{x_{k,\ell}}+\frac{\Delta x}{4}\zeta_{x_{k+1,\ell}}\right)-1,\notag\\
\textbf{E}_{2_{k,\ell}}&=2\pi
\left[\phi_{1,\ell}+\frac{\Delta x}{2}\phi_{x_{1,\ell}}+\Delta x\sum\limits_{i=2}^{k-1}\phi_{x_{i,\ell}}+\frac{3\Delta x}{4}\phi_{x_{k,\ell}}+\frac{\Delta x}{4}\phi_{x_{k+1,\ell}}-x^*_k\right]+\frac{\epsilon}{\left({x^*_k}^2+{y^*_\ell}^2+1 \right)^\frac{1}{2}}\notag\\
&-\sum\limits_{i=1}^{N}\sum\limits_{j=1}^{M} w(i,j)\left[\zeta_{x_{i,j}} -\frac{1}{2}(\zeta_{x_{k,\ell}}+\zeta_{x_{k+1,\ell}})\right]K_{3_{i,j,k,\ell}}-\frac{1}{2}(\zeta_{x_{k,\ell}}+\zeta_{x_{k+1,\ell}})I,\notag\\
\textbf{E}_{3_{\ell}}&=x_1\phi_{x_{1,\ell}}+n\phi_{1,\ell}-x_1(n+1),\label{eqn:SystemEquationsInZetaX}\\
\textbf{E}_{4_{\ell}}&=\frac{x_1}{\Delta x}\phi_{x_{2,\ell}}+(n-\frac{x_1}{\Delta x})\phi_{x_{1,\ell}}-n,\notag\\
\textbf{E}_{5_{\ell}}&=x_1\zeta_{x_{1,\ell}}+n\zeta_{1,\ell},\notag\\
\textbf{E}_{6_{\ell}}&=\frac{x_1}{\Delta x}\zeta_{x_{2,\ell}}+(n-\frac{x_1}{\Delta x})\zeta_{x_{1,\ell}},\notag
\end{align}
for $k=1\dots(N-1)$, $\ell=1\dots M$.

Finally, to calculate the linear Jacobian, the equations in (\ref{eqn:SystemEquationsInZetaX}) can be differentiated with respect to $\phi_{1,m}$, $\phi_{x_{n,m}}$, $\zeta_{1,m}$ and $\zeta_{x_{n,m}}$ to give:
\begin{align}
\frac{\partial \textbf{E}_{1_{k,\ell}}}{\partial \phi_{1,m}}&= 0,\notag\\
\frac{\partial \textbf{E}_{1_{k,\ell}}}{\partial \phi_{x_{n,m}}}&=
\begin{cases}
\frac{1}{2} & \text{for }n=k,k+1\text{ and }m=\ell\\
0 & \text{otherwise}\\
\end{cases},\label{eq:PreA}\\
\frac{\partial \textbf{E}_{1_{k,\ell}}}{\partial \zeta_{1,m}}&=
\begin{cases}
\frac{1}{F^2} & \text{for }m=\ell\\
0 & \text{otherwise}\\
\end{cases},\notag\\
\frac{\partial \textbf{E}_{1_{k,\ell}}}{\partial \zeta_{x_{n,m}}}&=
\begin{cases}
\frac{\Delta x}{4F^2} & \text{for }n=1\text{, }k=1\text{ and }m=\ell\\
\frac{\Delta x}{2F^2}& \text{for }n=1\text{, }k>1\text{ and }m=\ell\\
\frac{\Delta x}{F^2} & \text{for }n<k\text{ and }m=\ell\\
\frac{3\Delta x}{4F^2}& \text{for }n=k\text{ and }m=\ell\\
\frac{\Delta x}{4F^2} & \text{for }n=k+1\text{ and }m=\ell\\
0 & \text{otherwise}\\
\end{cases},\label{eq:PreB}\\
\frac{\partial \textbf{E}_{2_{k,\ell}}}{\partial \phi_{1,m}}&=
\begin{cases}
2\pi & \text{for }m=\ell\\
0 & \text{otherwise}\\
\end{cases},\notag\\
\frac{\partial \textbf{E}_{2_{k,\ell}}}{\partial \phi_{x_{n,m}}}&=
\begin{cases}
\frac{\pi\Delta x}{2} & \text{for }n=1\text{, }k=1\text{ and }m=\ell\\
\pi\Delta x & \text{for }n=1\text{, }k>1\text{ and }m=\ell\\
2\pi\Delta x & \text{for }n<k\text{ and }m=\ell\\
\frac{3\pi\Delta x}{2}& \text{for }n=k\text{ and }m=\ell\\
\frac{\pi\Delta x}{2} & \text{for }n=k+1\text{ and }m=\ell\\
0 & \text{otherwise}\\
\end{cases},\label{eq:PreC}
\end{align}
\begin{align}
\frac{\partial \textbf{E}_{2_{k,\ell}}}{\partial \zeta_{1,m}}&= 0\notag\\
\frac{\partial \textbf{E}_{2_{k,\ell}}}{\partial \zeta_{x_{n,m}}}&=
\begin{cases}
\frac{1}{2}\sum\limits_{i=1}^{N}\sum\limits_{j=1}^{M}w(i,j)K_{3_{i,j,k,\ell}}-\frac{1}{2}I-w(n,m)K_{3_{n,m,k,\ell}} & \text{for }n=k,k+1\text{ and }m=\ell\\
-w(n,m)K_{3_{n,m,k,\ell}} & \text{otherwise}\\
\end{cases}\label{eq:PreD},
\end{align}
for $k=1\dots(N-1)$, $\ell=1\dots M$ and $n=1\dots N$, $m=1\dots M$. The derivatives for $\textbf{E}_{3_{\ell}},\textbf{E}_{4_{\ell}},\textbf{E}_{5_{\ell}},\textbf{E}_{6_{\ell}}$ can be easily calculated and will not be explicitly written here. Our preconditioner is formed by ordering these Jacobian entries in the manner described in \ref{appendixA}.

\section{Preconditioner storage and factorisation}\label{appendixC}
\label{sec:InvPre}
As shown in Figure~\ref{fig:TwoJacVis}, the preconditioner can be divided up into four equal submatrices of size $(N+1)M\times(N+1)M$. This preconditioner can then be factorised using the block decomposition,
\[
\textbf{P}=\left[
\begin{matrix}
A & B\\
C & D\\
\end{matrix}\right]=
\left[\begin{matrix}
I & 0\\
CA^{-1} & I\\
\end{matrix}\right]
\left[\begin{matrix}
A & 0\\
0 & D-CA^{-1}B\\
\end{matrix}\right]
\left[\begin{matrix}
I & A^{-1}B\\
0 & I\\
\end{matrix}\right],
\]
where $A$, $B$, $C$, and $D$ are primarily given by equations (\ref{eq:PreA}), (\ref{eq:PreB}), (\ref{eq:PreC}) and (\ref{eq:PreD}), respectively. Thus we can solve the system $\textbf{P}\textbf{r}=\textbf{b}$ by performing the following operations,
\begin{equation*}
\left[\begin{matrix}
\textbf{t}_1\\
\textbf{t}_2
\end{matrix}\right]
=
\left[\begin{matrix}
\textbf{b}_1\\
\textbf{b}_2-CA^{-1}\textbf{b}_1
\end{matrix}\right],\quad
\left[\begin{matrix}
\textbf{s}_1\\
\textbf{s}_2
\end{matrix}\right]
=
\left[\begin{matrix}
A^{-1}\textbf{t}_1\\
(D-CA^{-1}B)^{-1}\textbf{t}_2
\end{matrix}\right],\quad
\left[\begin{matrix}
\textbf{r}_1\\
\textbf{r}_2
\end{matrix}\right]
=
\left[\begin{matrix}
\textbf{s}_1-A^{-1}B\textbf{s}_2\\
\textbf{s}_2
\end{matrix}\right].
\end{equation*}
This method provides several advantages. First, $A$ is tridiagonal, allowing for easy storage and fast factorisation and inversion when needed.  Second, $B$ and $C$ are only used in matrix vector multiplication operations and thus can be implemented as functions that perform these operations rather than stored as matrices. Furthermore, $A$, $B$ and $C$ are block diagonal, and each diagonal block is identical within a given matrix, meaning $CA^{-1}B$ need only be computed for one block.  Finally, $D$ appears only in the Schur complement $(D-CA^{-1}B)$, which we store and factorise in the preconditioner set-up phase.  These advantages mean we only store a $3\times(N+1)M$ matrix for $A$ and a block-banded matrix for the Schur complement $(D-CA^{-1}B)$ when constructing and factorising the preconditioner.

\end{document}